\documentclass[prd,aps,tightenlines,floats,nofootinbib,preprintnumbers,superscriptaddress]{revtex4-2}

\usepackage{graphicx}
\usepackage[caption=false]{subfig}
\usepackage{amsmath, amssymb, xcolor, bm}
\usepackage[colorlinks=true,linkcolor=black,citecolor=olive,urlcolor=blue]{hyperref}

\begin{document}

\title{Gravitational Waves from Superconducting Cosmic Strings}

\author{Jinyoung Jhun}
\email{j.jhun@rikkyo.ac.jp} 
\affiliation{Department of Physics, Rikkyo University, Toshima, Tokyo 171-8501, Japan}

\author{Takashi Hiramatsu}
\email{hiramatsu.takashi@nihon-u.ac.jp}
\affiliation{Department of Physics, Rikkyo University, Toshima, Tokyo 171-8501, Japan}
\affiliation{Department of Physics, College of Science and Technology, Nihon University, 1-8-14 Kanda-Surugadai, Chiyoda-ku, Tokyo 101-8308, Japan}

\preprint{RUP-26-17}

\begin{abstract}
We study the evolution of superconducting cosmic-string networks in a $U(1)_{\rm local}\times U(1)_{\rm global}$ field-theoretic model using three-dimensional lattice field simulations in an expanding universe. The scalar field charged under $U(1)_{\rm local}$ forms cosmic strings, while the scalar field associated with $U(1)_{\rm global}$ acts as a current carrier and condenses in their vicinity. We investigate the evolution of the string network for several values of the coupling constant between the string-forming and current-carrier fields and estimate the gravitational-wave power spectrum sourced by the resulting superconducting cosmic-string network. We find that the spectral shape depends on the coupling constant: as the interaction strength increases, the spectrum is suppressed on small scales. This feature may be probed by future high-frequency gravitational-wave observations.
\end{abstract}

\maketitle

%======================================================================%
\section{Introduction}
%======================================================================%

Many hypotheses beyond the Standard Model, such as grand unified theories, involve spontaneous symmetry breaking from a larger symmetry group to that of the Standard Model at low energies. In the early universe, such symmetry breaking is expected to occur through phase transitions induced by the temperature decrease associated with cosmic expansion. During these phase transitions, the fields settle onto the vacuum manifold, or equivalently the order-parameter space, determined by the symmetry-breaking pattern. If the topology of the vacuum manifold is nontrivial, topological defects can form through the Kibble–Zurek mechanism \cite{Kibble:1976sj,Zurek:1985qw}. In particular, when the vacuum manifold has the topology of $S^1$, one-dimensional topological defects, known as cosmic strings, can be produced. (See Refs.~\cite{Vilenkin:2000jqa, Manton:2004tk, Weinberg:2012pjx} for topological defects associated with other vacuum-manifold topologies.)

The presence of a topological defect can also trigger symmetry breaking in an additional field sector \cite{Montonen:1976yk,Sarkar:1976vr,Witten:1984eb,Peter:1995ks,2005IJMPB..19.1835K,Lilley:2010av,Nitta:2013wca}. A well-known cosmological example is provided by superconducting cosmic strings, originally proposed by Witten \cite{Witten:1984eb}. In this framework, an additional scalar field condenses inside the string core, thereby breaking an additional $U(1)$ symmetry and rendering the string current-carrying. This mechanism is closely analogous to that underlying the Ginzburg--Landau description of superconductivity \cite{Ginzburg:1950sr}, which is the origin of the name ``superconducting cosmic strings.'' 
Superconducting strings and their physical properties have been studied extensively in the literature \cite{
Copeland:1987th, Babul:1987me, Everett:1988tn, Peter:1992dw, Peter:1992ta, Peter:1993tm, Larsen:1993ha, Davis:1995kk, Allen:1995rd, Davis:1996xs, Kibble:1996rh, Dimopoulos:1997xa, Martins:1997nb, Martin:1998cv, Dimopoulos:1999dn, Blanco-Pillado:2002vwq, Garaud:2009uy, Lilley:2009yr, Lilley:2010av, Oikonomou:2010ut, Babeanu:2011ie, Hartmann:2016axn, Hartmann:2017lno, Fukuda:2020kym, Abe:2020ure}. The presence of a current supported by a condensate localized on the
string can give rise to distinctive physical effects compared with
non-current-carrying cosmic strings, such as electromagnetic radiation~\cite{Vilenkin:1986zz,Spergel:1986uu, Garfinkle:1987yw, Copeland:1987yv,Garfinkle:1988yi,Blanco-Pillado:2000nbp, Blanco-Pillado:2000ssd, Ferrer:2005xva,Cai:2012zd, Miyamoto:2012ck,Tashiro:2012nv, Imtiaz:2020igv}, cosmic-ray emission \cite{Berezinsky:2009xf,Vachaspati:2009kq}, the global 21-cm signal \cite{Theriault:2021mrq,Cyr:2023iwu,Cyr:2023yvj,Si:2025vsj}, and dark photon production \cite{Long:2019lwl}.

The presence of a current also affects gravitational waves emitted by cosmic string networks. Several analytical studies of gravitational radiation from superconducting cosmic strings have been carried out using models in which the cosmic string is described by the effective worldsheet Lagrangian, an extension of the Nambu--Goto string description, with both the string itself and the current flowing along it treated as infinitely thin \cite{Rybak:2022sbo, Rybak:2024our}. Furthermore, in order to evaluate the stochastic gravitational-wave background, it is also necessary to take into account the evolution of the cosmic string network. For superconducting cosmic strings, the charge-velocity-dependent one-scale model \cite{Martins:2020jbq, Martins:2021cid}, which extends the velocity-dependent one-scale model by incorporating the macroscopic effects of the current and charge, has been employed in several studies of network evolution \cite{Rybak:2022sbo, Auclair:2022ylu}.

The conventional worldsheet effective description of cosmic strings provides a good approximation when inter-string interactions are negligible. In general, however, cosmic strings can interact with one another \cite{Fujikura:2023lil}, and the evolution of a string network is known to depend sensitively on these interactions \cite{Hiramatsu:2013tga,Zhao:2025kvm}.
A representative example is provided by the Abelian-Higgs model, one of the standard field-theoretic models giving rise to cosmic strings. In this model, the relative strength of the gauge and scalar interactions determines whether the effective force between strings is attractive or repulsive, corresponding to the type-I and type-II regimes, respectively. Several network simulations have demonstrated that this distinction leads to differences in network evolution \cite{Hiramatsu:2013tga,Hindmarsh:2018wkp}. Importantly, such interaction-dependent effects have been shown to have a non-negligible impact on observational constraints from the Cosmic Microwave Background \cite{Hindmarsh:2018wkp}.
Superconducting cosmic strings also possess inter-string interactions \cite{Abe:2020ure, Fujikura:2023lil}, and their effects may therefore be non-negligible in the evolution of the network.

In this paper, we study the evolution of superconducting cosmic string networks
in a toy model with $U(1)_{\rm local}\times U(1)_{\rm global}$ symmetry by means
of three-dimensional lattice field simulations. The purpose of this work is to
clarify how the current-carrier condensate and its interaction with the
string-forming field affect both the network evolution and the gravitational-wave spectrum. Unlike previous lattice studies of superconducting cosmic strings
\cite{Correia:2024wsq}, which employed the fat-string regime, we perform the
simulations in the physical-string regime, where the comoving string width decreases with cosmic expansion. This setup allows us to examine whether currents and charges are generated on the strings under more realistic string-width evolution and how their presence modifies the gravitational-wave signal from the network. We show that the coupling constant between the string-forming and current-carrier fields changes the shape of the gravitational-wave power spectrum, and we identify the respective roles of the string and current-carrier field in this modification. We also discuss the observational implications of the resulting spectrum and show that the characteristic differences from Abelian-Higgs cosmic strings are expected to appear at frequencies in the GHz range or above.

This paper is organized as follows. In Sec.~II, we present the model we use in the simulation and derive the field equations in a radiation-dominated background. In Sec.~III, we introduce the numerical setup for the simulation, and in Sec.~IV, we give the numerical results. Finally, we present our conclusions and discussion in Sec.~V.

%======================================================================%
\section{Basic equations}
We consider an action
\begin{align}
    S_M[\Phi,\sigma,A_\mu,g_{\mu\nu}] &= -\int\!d^4x\,\sqrt{-g}\left[ (D_\mu\Phi)^*(D^\mu\Phi) + (\partial_\mu\sigma)^*(\partial^\mu\sigma) + \frac{1}{4}F_{\mu\nu}F^{\mu\nu} + U(\Phi,\sigma)\right],
    \label{eq:action}\\
    \quad \label{potential}U(\Phi,\sigma) &= \frac{\lambda_\Phi}{4}(|\Phi|^2-\eta^2)^2 + \lambda_{\Phi\sigma}(|\Phi|^2-\eta^2)|\sigma|^2 + \frac{m_\sigma^2}
{2}|\sigma|^2 + \frac{\lambda_\sigma}{4}|\sigma|^4,
\end{align}
which has $U(1)_{\text{local}}\times U(1)_{\text{global}}$ symmetry \cite{Peter:1992dw}, where $\Phi$ is a complex scalar field charged under the local $U(1)$ symmetry, coupled to the gauge field $A_\mu$ through the covariant derivative $D_\mu = \partial_\mu - ieA_\mu$, with the field strength defined as $F_{\mu\nu} \equiv \partial_\mu A_\nu - \partial_\nu A_\mu $.
The spontaneous symmetry breaking of $U(1)_{\rm local}$ may yield cosmic strings.

The current-carrier field $\sigma$ is a complex scalar field with the global $U(1)$ symmetry, whose condensation
on the strings renders them superconducting.
The parameter regions that admit stable, straight, and static superconducting cosmic string configurations have been studied in Ref.~\cite{Hiramatsu:2023epr}.

We vary the action and obtain the equations of motion for $\Phi$, $\sigma$, and $A_{\mu}$ with $g_{\mu\nu}=a^2\eta_{\mu\nu}$, where $a$ is the scale factor and $\eta_{\mu\nu}=\mathrm{diag}(-1,1,1,1)$ is the Minkowski metric,
\begin{align}
    \begin{split}\Phi'' + 2\mathcal{H}\Phi' -\Delta\Phi &+ ie\eta^{\mu\nu}(\partial_\mu A_\nu)\Phi + 2ie\eta^{\mu\nu}A_\mu\partial_\nu\Phi -2ie\mathcal{H}A_0\Phi + e^2\eta^{\mu\nu}A_\mu A_\nu\Phi\\
    &= -a^2\left[ \frac{\lambda_{\Phi}}{2}(|\Phi|^2-\eta^2)\Phi + \lambda_{\Phi\sigma}|\sigma|^2\Phi \right],\end{split} \label{eq:Phi} \\
    \sigma'' + 2\mathcal{H}\sigma' -\Delta\sigma &= -a^2\left[ \lambda_{\Phi\sigma}(|\Phi|^2-\eta^2)\sigma + \frac{m^2_\sigma}{2}\sigma + \frac{\lambda_\sigma}{2}|\sigma|^2\sigma \right], \label{eq:sigma}\\
    A''_\alpha - \Delta A_\alpha + \partial_\alpha(\eta^{\mu\nu}\partial_\mu A_\nu) &= a^2[2e\;\text{Im}(\Phi^*\partial_\alpha\Phi) - 2e^2A_\alpha|\Phi|^2], \label{eq:A}
\end{align}
where the prime denotes the derivative with respect to the conformal time and $\mathcal{H}\equiv aH$ is the conformal Hubble parameter. Throughout the paper, we assume a radiation-dominated universe and choose the temporal gauge, $A_0=0$.
From Eq.~(\ref{eq:A}) with $\alpha=0$, we obtain a constraint equation,
\begin{align}
    \delta^{ij}\partial_iA'_j = 2ea^2\;\text{Im}(\Phi^*\Phi').
     \label{eq:constraint}
\end{align}

Additionally, we consider the tensor perturbations of the spacetime metric, $g_{\mu\nu}=a^2(\eta_{\mu\nu}+h_{\mu\nu})$. At linear order, the governing equation is given by
\begin{align}
h_{ij}'' + 2\mathcal{H}h_{ij}' - \Delta h_{ij} = 2 M_{\rm Pl}^{-2} T_{ij},
\end{align}
where $M_{\rm Pl} \equiv 1/\sqrt{8\pi G}$ is the reduced Planck mass. The right-hand side is the energy-momentum tensor of the scalar fields and the gauge field given by the variation of the action (\ref{eq:action}) with respect to $g^{\mu\nu}$,
\begin{align}
T_{\mu\nu} = -\frac{2}{\sqrt{-g}}\frac{\delta S_M}{\delta g^{\mu\nu}} = 2\text{Re}[(D_{\mu}\Phi)^* D_{\nu}\Phi] + 2\text{Re}[(\partial_{\mu}\sigma)^* \partial_{\nu}\sigma] + F_{\mu \sigma}F_{\nu}{}^\sigma + g_{\mu\nu}\mathcal{L},
\end{align}
where $\mathcal{L}$ is the Lagrangian density appearing in the action. We retain only the relevant terms that source gravitational waves, which are given as
\begin{align}
T_{ij} \supset 
T_{ij}^{\text{eff}} = 2\text{Re}[(D_{i}\Phi)^* D_{j}\Phi] + 2\text{Re}[(\partial_{i}\sigma)^* \partial_{j}\sigma] + F_{i\mu}F_{j}{}^\mu. \label{eq:relavantEMT}
\end{align}
Therefore, in practice, we solve the following equation,
\begin{align}
h_{ij}'' + 2\mathcal{H}h_{ij}' - \Delta h_{ij} = 2 M_{\rm Pl}^{-2} T_{ij}^{\text{eff}}.
\label{eq:h}
\end{align}
The procedure for extracting the gravitational-wave power spectrum will be presented in Appendix \ref{sec:GWpower}.

%======================================================================%

%======================================================================%
\section{Numerical setup}
In numerical simulations, we solve the field equations in terms of the following rescaled fields and coordinates.
\begin{align}
\label{rescaled}
    \tilde\Phi \equiv \frac{\Phi}{\eta},\quad\tilde\sigma \equiv \frac{\sigma}{\eta},\quad\tilde A_i \equiv \frac{
A_i}{\eta},\quad\tilde x^\mu \equiv \eta x^\mu.
\end{align}
The initial conditions for the scalar fields are generated according to the power spectrum
\begin{align}
    P_I(k) \equiv \left\langle |Z_I(\boldsymbol{k})|^2 \right\rangle
    = A \exp\left(-\frac{k^2}{k_{\mathrm{cut}}^2}\right),
    \label{eq:initial}
\end{align}
where \(\boldsymbol{k}\) is the conformal wave vector, \(k=|\boldsymbol{k}|\), and
\(Z_I=(\mathrm{Re}\,\Phi,\;\mathrm{Im}\,\Phi,\;\mathrm{Re}\,\sigma,\;\mathrm{Im}\,\sigma)\).
${Z}_I({\boldsymbol{k}})$ denotes the Fourier coefficient of $Z_I(x)$, defined as
\begin{align}
Z_I({\boldsymbol{k}}) \equiv \frac{1}{\sqrt{V}}\int_V d^3 x Z_I( x) e^{-i{\boldsymbol{k}}\cdot{\boldsymbol{x}}},
\end{align}
and is taken to be a Gaussian random variable. Here, $V$ is the conformal volume of the simulation box, while $A$ is a parameter chosen by hand. A cutoff $k_\text{cut}$ is introduced to suppress unphysical behavior arising from ultraviolet modes.

For the other fields, we impose the following initial conditions:
\begin{align}
    A_i = h_{ij} = 0.
\end{align}
In addition, the first time derivatives of all fields are initially set to zero. The above initial conditions automatically satisfy the constraint equation, Eq.~(\ref{eq:constraint}), at the initial time. The scale factor is normalized as
\begin{align}
a(\tau) = \frac{\tau}{\tau_{\rm i}}, \qquad a(\tau_{\rm i}) = 1,
\end{align}
where $\tau_{\rm i}$ denotes the initial conformal time, which is fixed to $\tau_{\rm i}=1/\eta$ in this paper.

The conformal Hubble parameter can then be written as $\mathcal{H}=\tau^{-1}$. In this simulation, instead of specifying the comoving box size $L$ directly, we use the following quantity as a parameter, which represents how many horizon-sized boxes are contained within the simulation box:
\begin{align}
\frac{L}{\mathcal{H}^{-1}} = \frac{L}{\tau}.
\end{align}

We solve the equations of motion in a comoving simulation box whose physical side length expands with the cosmic scale factor in the radiation-dominated era. The final scale factor is fixed to $a_{\rm f}=16$. The numerical parameters used in our simulations are summarized in Table~\ref{tab:parameters}.

\begin{table*}[htbp]
\centering
\begin{tabular}{c|c|c|c|c|c|c|c|c|c|c|c|c}
      Set & $e$ &$\lambda_{\Phi}$ & $\lambda_{\sigma}$ & $\lambda_{\Phi\sigma}$ & ${m_\sigma^2}/2\eta^2$ & $A \eta$ & $k_{\text{cut}}/\eta$ & $L/{\mathcal{H}_{\rm i}^{-1}}$ & $L/{\mathcal{H}_{\rm f}^{-1}}$ & $\eta\Delta x_{\rm phys,f}$ & $N^3$ & $N_s$\\ \hline \hline
        1 &$\frac{1}{\sqrt{2}}$ &1.0 & 1.0 & {\rm various} & 0.01   & $1.0\times 10^{-4}$ & 10.0 & $32$ & $2$& $0.5$ & $1024^3$ & 2560\\  \hline
        2 &$\frac{1}{\sqrt{2}}$ &1.0 & 1.0 & 0.49 & 0.01   & $1.0\times 10^{-4}$ & 10.0 & $32$ & $2$& $0.25$& $2048^3$ & 5120\\  \hline
        3 &$\frac{1}{\sqrt{2}}$ &1.0 & 1.0 & 0.49 & 0.01   & $1.0\times 10^{-4}$ & 10.0 & $64$ & $4$& $0.5$ & $2048^3$ & 2560
\end{tabular}
\caption{Numerical parameters used in the simulations. Here, $e$, $\lambda_{\Phi}$, $\lambda_{\sigma}$, and $m_\sigma$ are the parameters appearing in the potential~(\ref{potential}); $A$ and $k_{\rm cut}$ are the amplitude and cutoff scale, respectively, of the initial power spectrum introduced in Eq.~(\ref{eq:initial}); $L/{\mathcal{H}_{\rm i}^{-1}}$ and $L/{\mathcal{H}_{\rm f}^{-1}}$ are the initial and final ratios, respectively, of the comoving box size to the horizon size; $\eta\Delta x_{\rm phys,f}$ is the dimensionless physical lattice spacing at the final time; $N^3$ denotes the number of lattice points; and $N_s$ denotes the number of time steps.
}
\label{tab:parameters}
\end{table*}

Table~\ref{tab:parameters} lists the model parameters used in the simulations. In Set 1, $\lambda_{\Phi\sigma}$ is varied up to $0.49$ to study its impact on the dynamics and the resulting gravitational-wave power
spectrum. 
The upper end of this range is motivated by Ref.~\cite{Hiramatsu:2023epr}, where the corresponding coupling was shown to yield a superconducting string with the current-carrier field clearly localized in its core. 
In Sets 2 and 3, we fix $\lambda_{\Phi\sigma}=0.49$ to examine the dependence of the gravitational-wave power spectrum on the lattice resolution and the box size (see Fig.~\ref{fig:Omegaboxdxdep}). The convergence check is performed by varying the initial and final ratios of the comoving box size to the horizon size, $L/{\mathcal{H}_{\rm i}^{-1}}$ and $L/{\mathcal{H}_{\rm f}^{-1}}$, the physical lattice spacing at the final time, $\Delta x_{\rm phys,f}$, and the number of lattice points, $N^3$, while keeping the physical setup unchanged. Compared with Set 1, Set 2 employs a factor-of-two finer lattice spacing, whereas Set 3 uses twice the comoving box size.

%======================================================================%

%======================================================================%
\section{Results}
%======================================================================%

%-----------------------------------------------------------%
\subsection{String network}
%-----------------------------------------------------------%

\begin{figure}[htbp]
    \centering

    \begin{minipage}{\linewidth}
      \begin{minipage}{0.32\linewidth}
         \centering
         $\tau = 8.5\tau_i$
      \end{minipage}
      \begin{minipage}{0.32\linewidth}
         \centering
         $\tau = 12.25\tau_i$
      \end{minipage}
      \begin{minipage}{0.32\linewidth}
         \centering
         $\tau = 16\tau_i$
      \end{minipage}
    \end{minipage}

    \begin{minipage}{\linewidth}
        \centering
        \includegraphics[height=5.4cm,keepaspectratio]
          {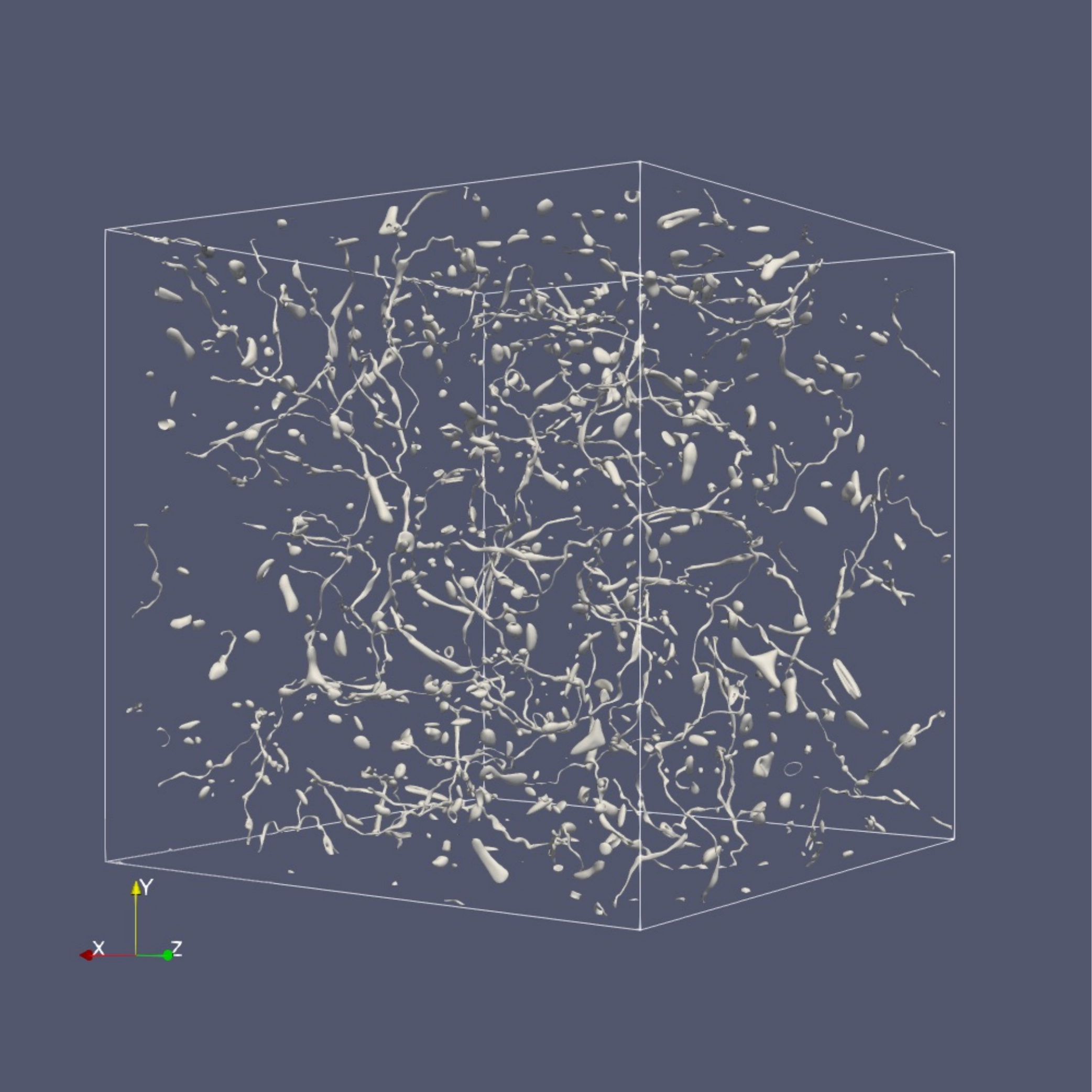}
        \includegraphics[height=5.4cm,keepaspectratio]
          {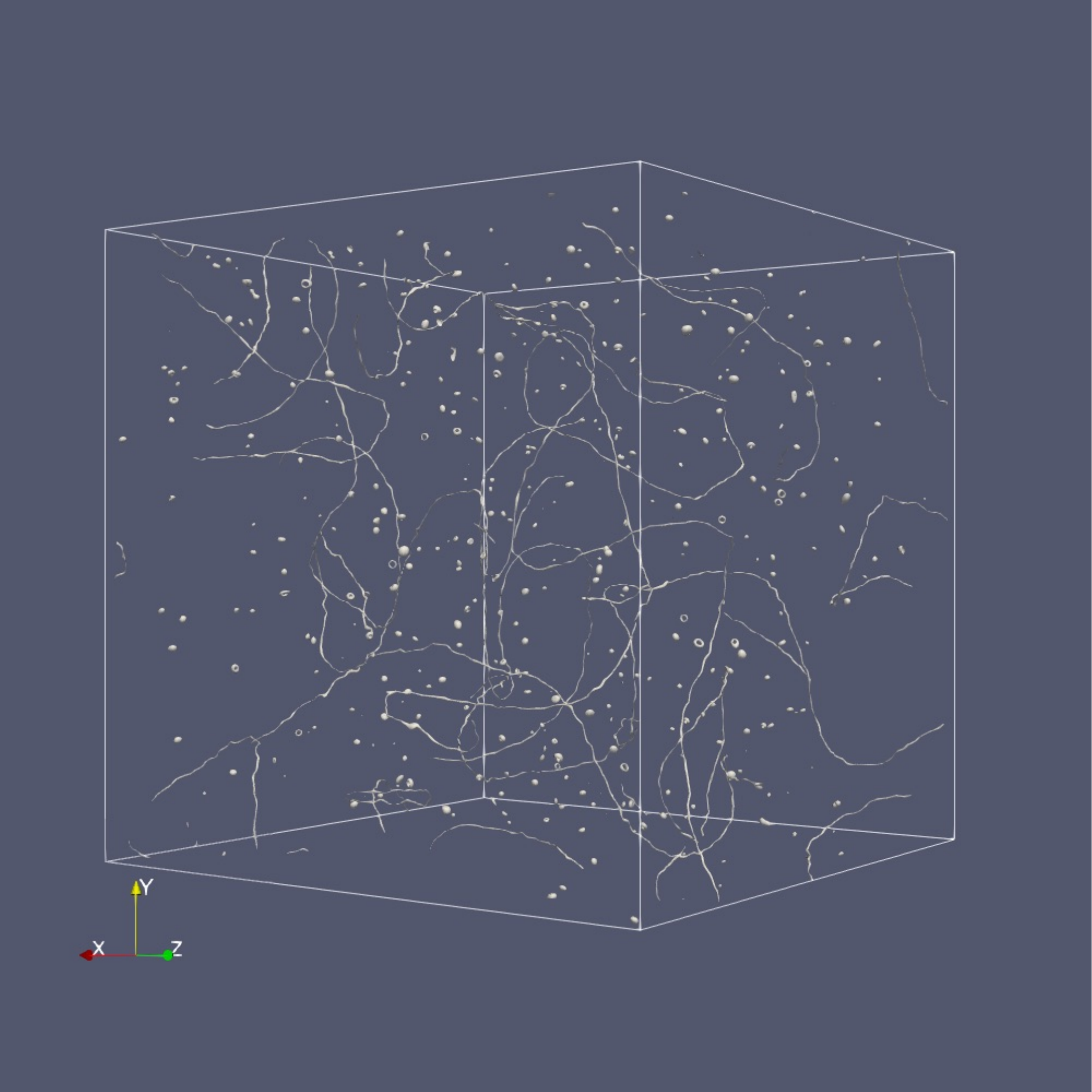}
        \includegraphics[height=5.4cm,keepaspectratio]
          {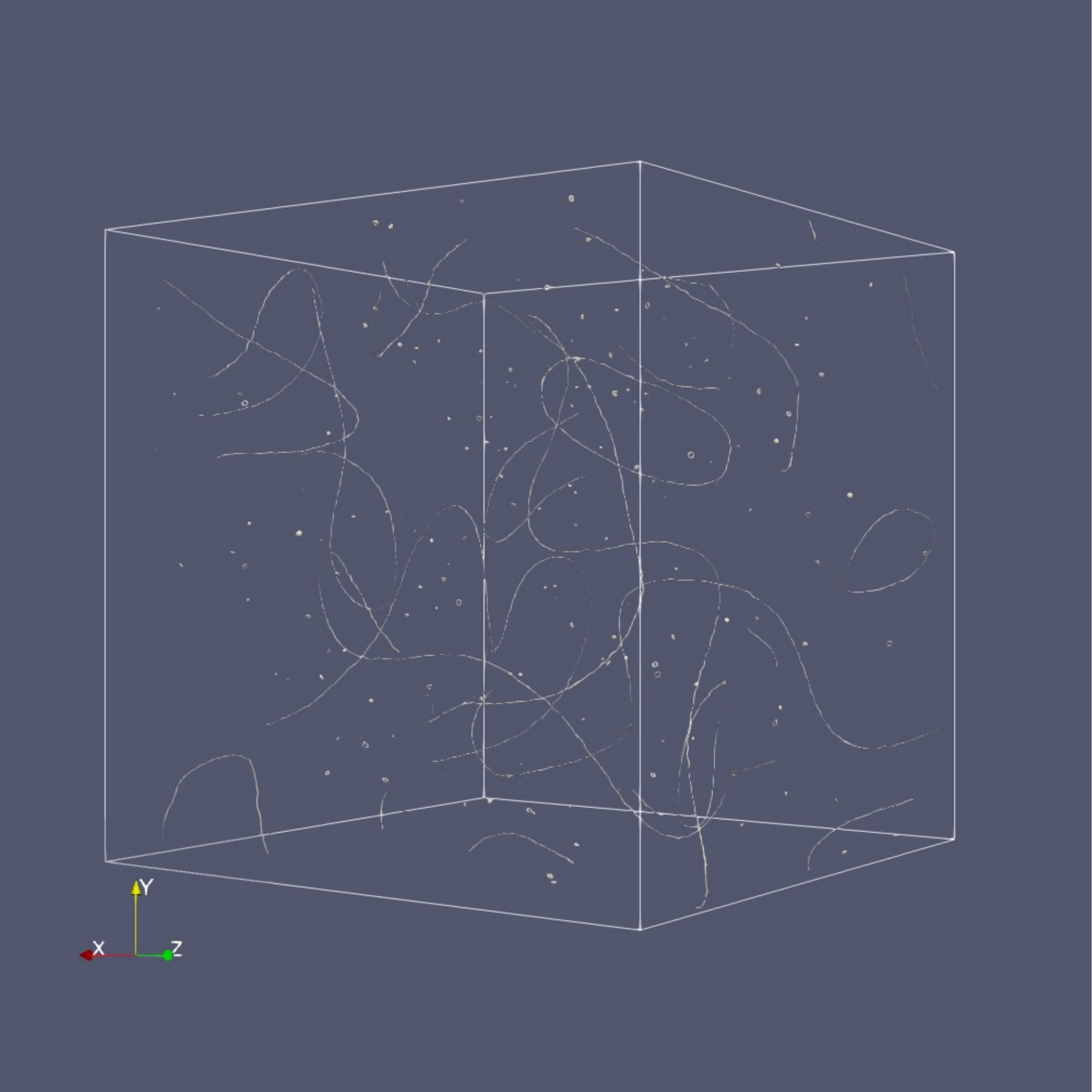}

        \par
        $\lambda_{\Phi\sigma}=0.00$
        \par
    \end{minipage}

    \begin{minipage}{\linewidth}
        \centering
        \includegraphics[height=5.4cm,keepaspectratio]
          {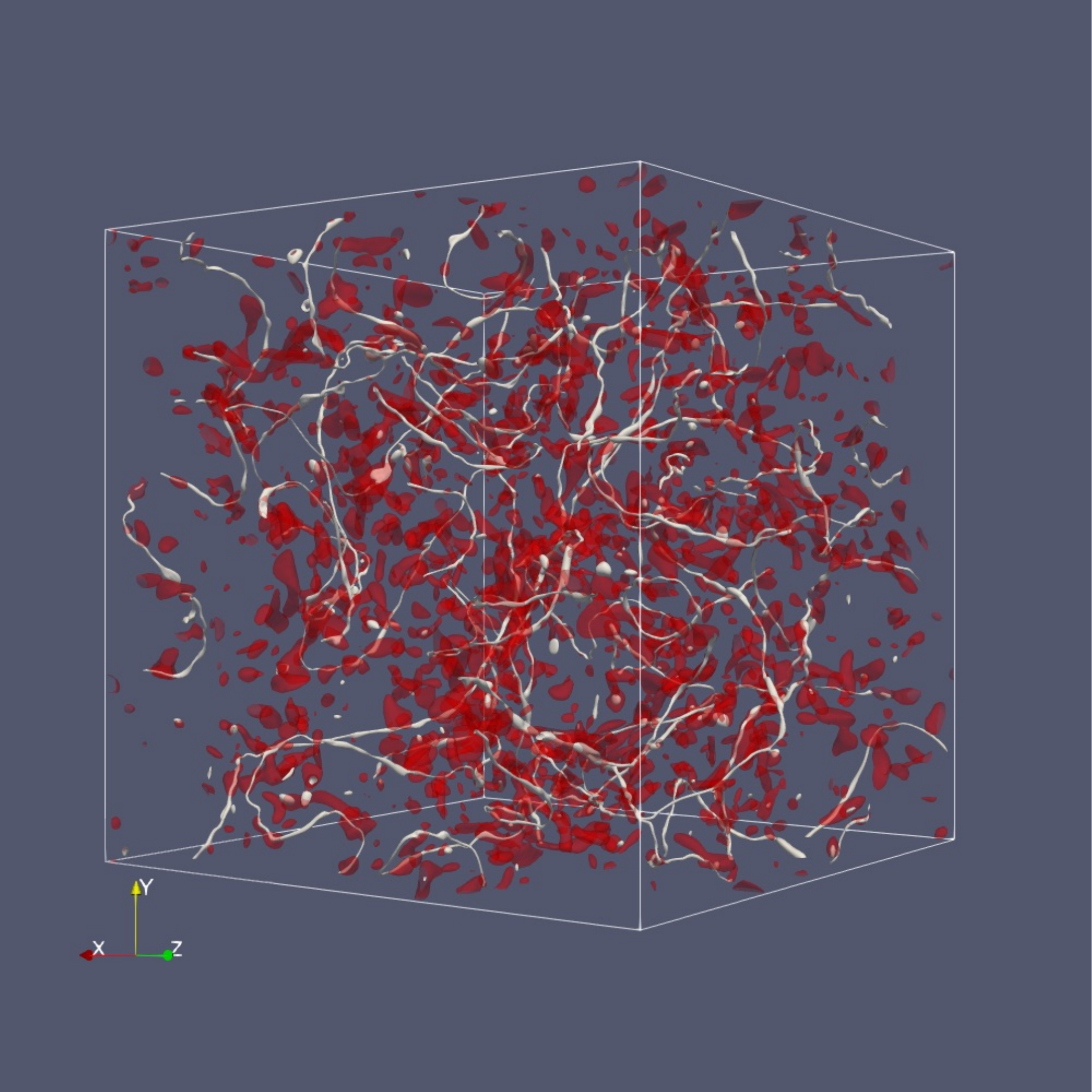}
        \includegraphics[height=5.4cm,keepaspectratio]
          {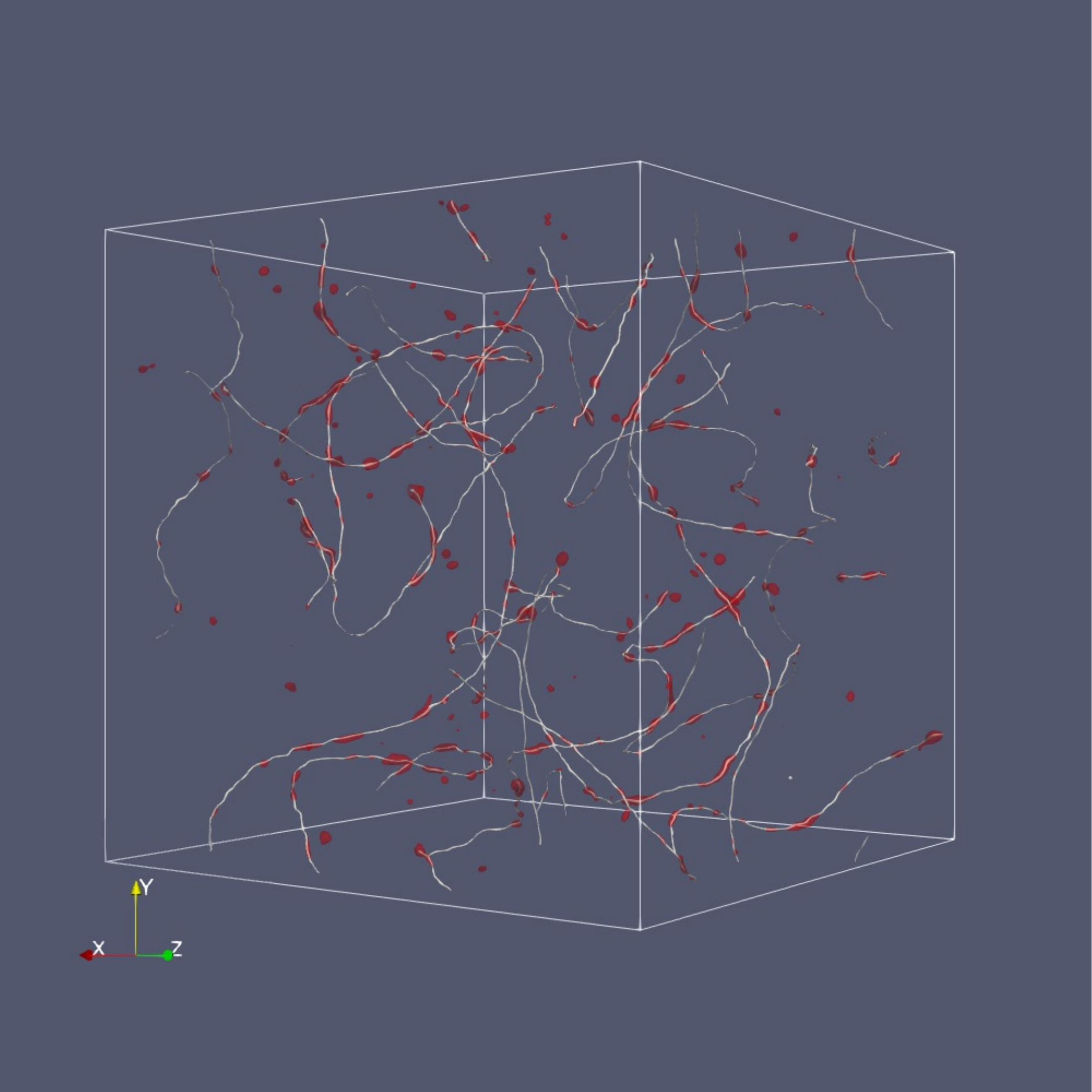}
        \includegraphics[height=5.4cm,keepaspectratio]
          {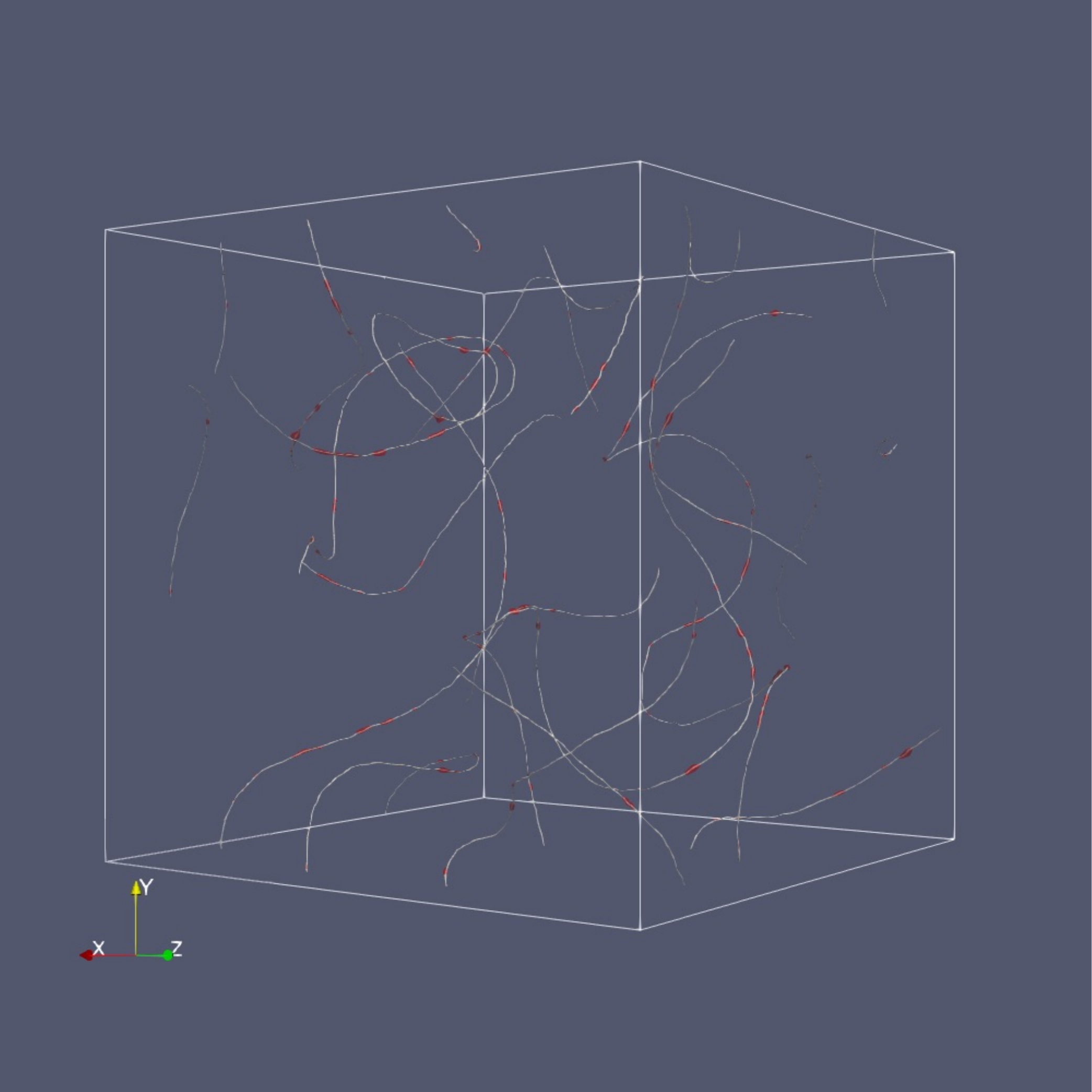}

        \par
        $\lambda_{\Phi\sigma}=0.40$
        \par
    \end{minipage}

    \begin{minipage}{\linewidth}
        \centering
        \includegraphics[height=5.4cm,keepaspectratio]
          {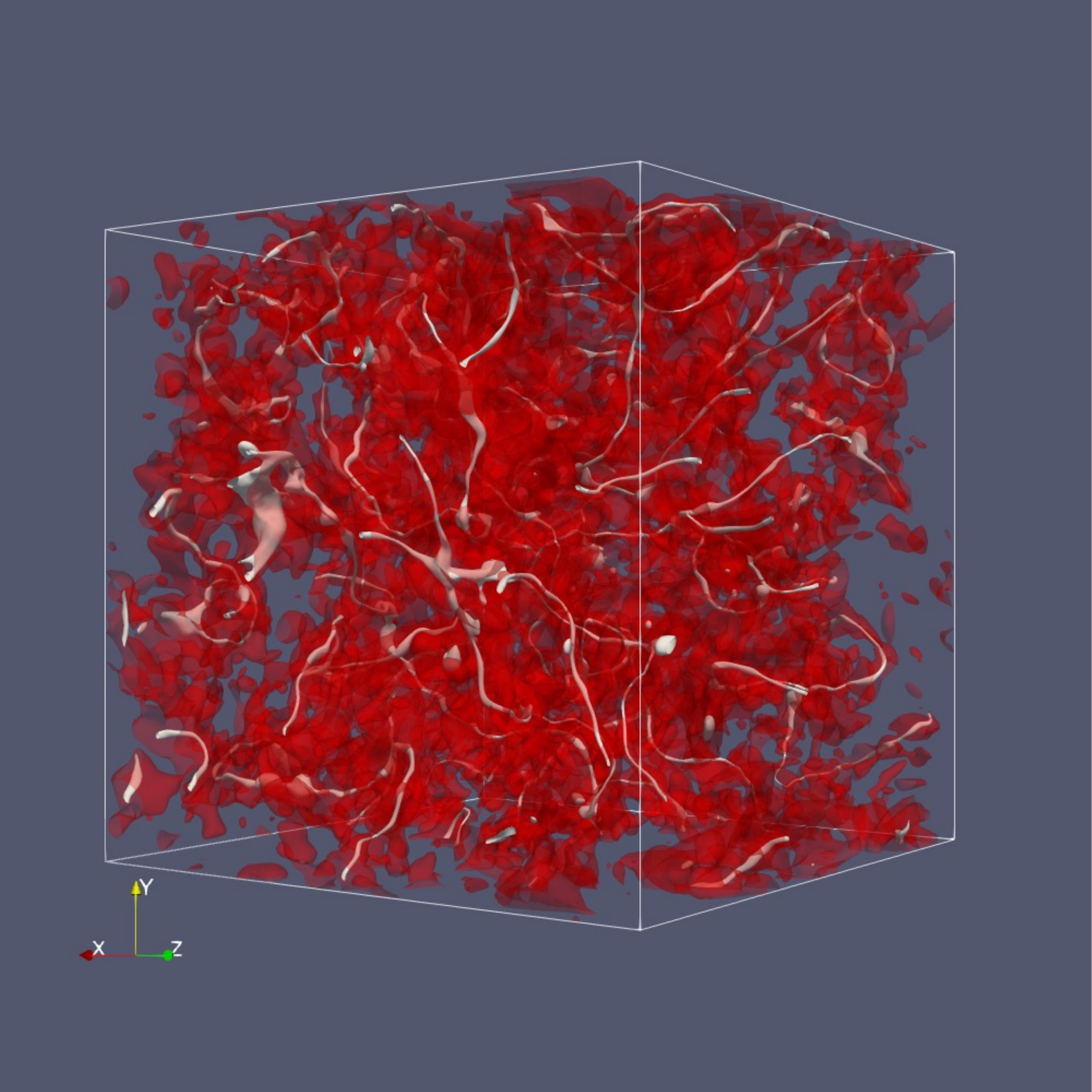}
        \includegraphics[height=5.4cm,keepaspectratio]
          {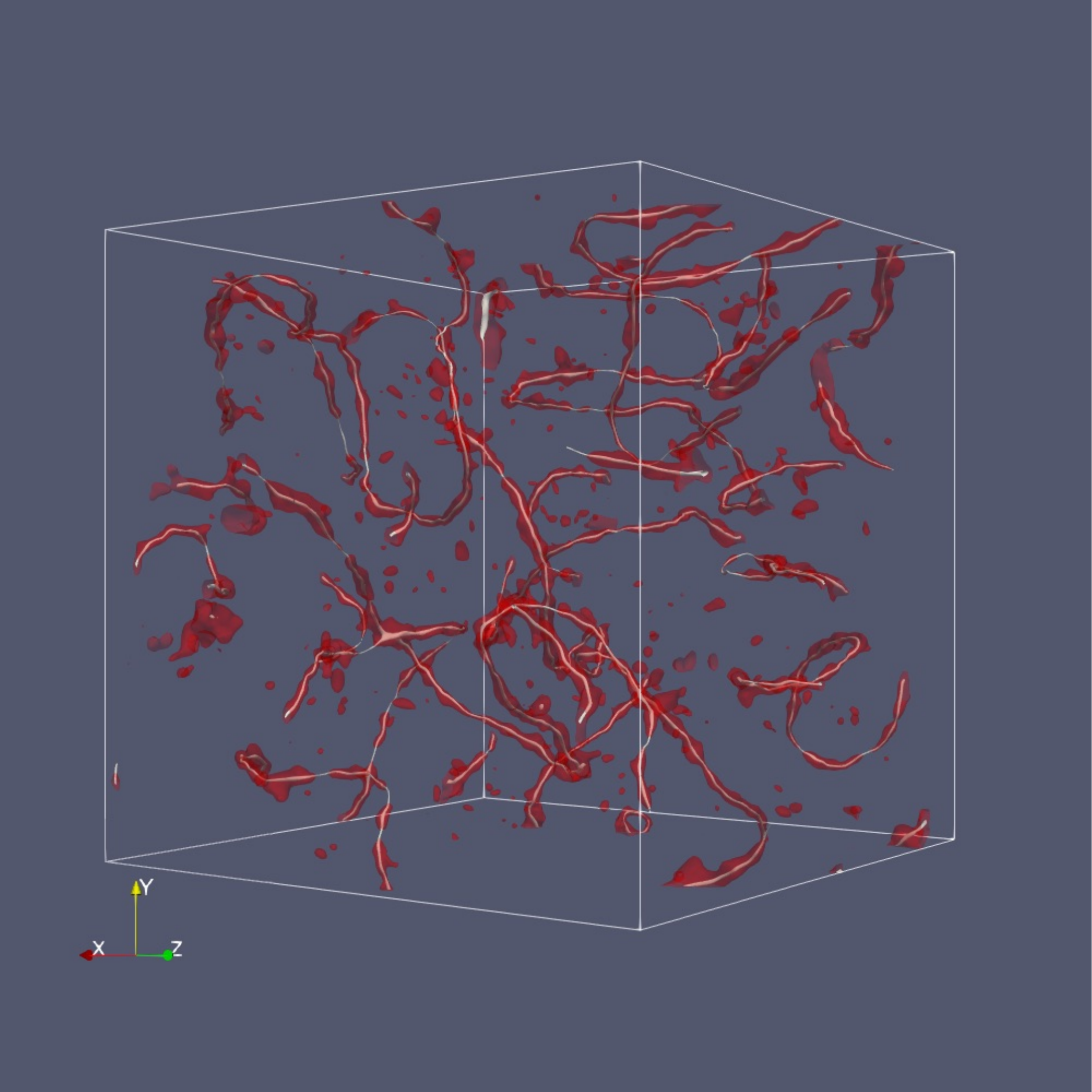}
        \includegraphics[height=5.4cm,keepaspectratio]
          {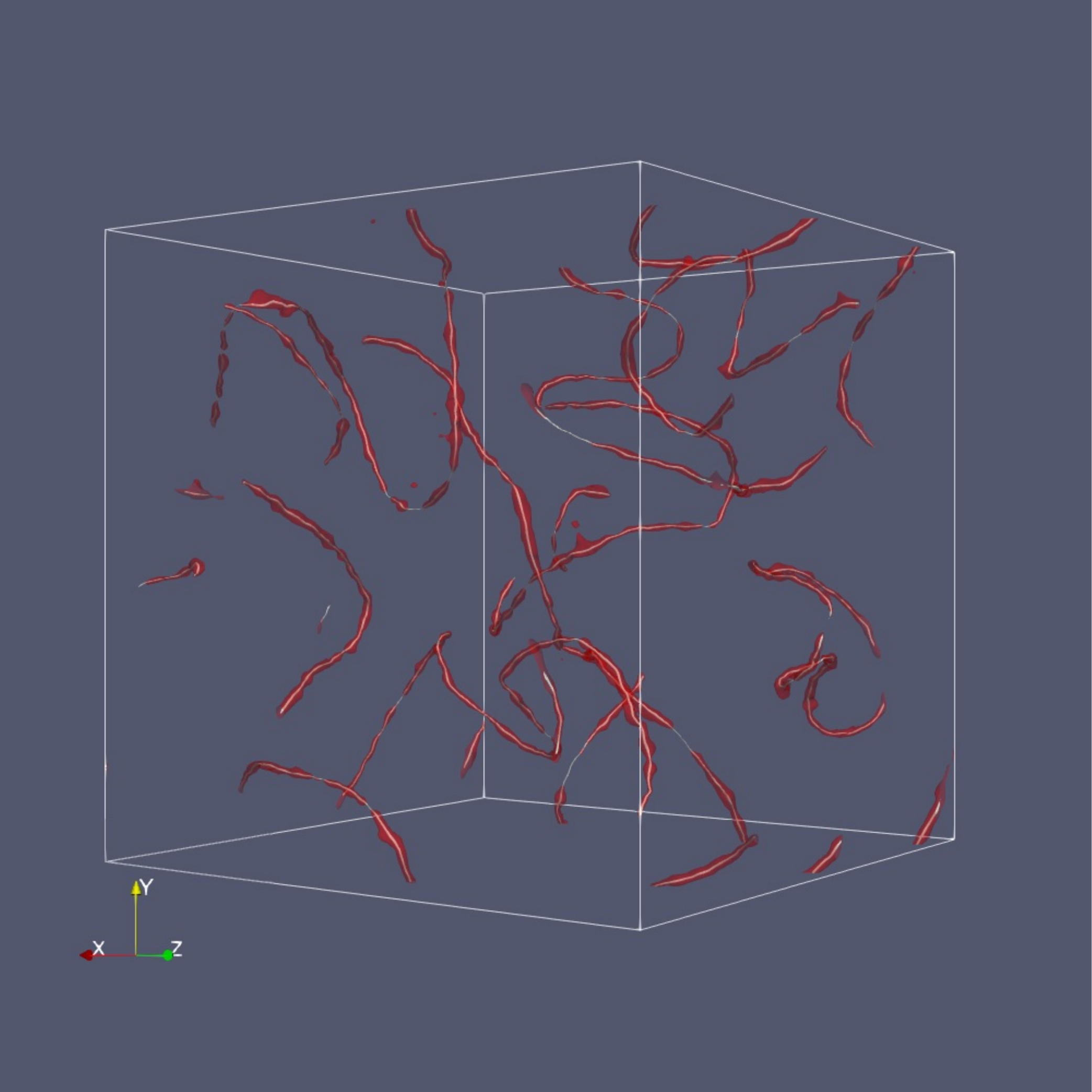}

        \par
        $\lambda_{\Phi\sigma}=0.49$
        \par
    \end{minipage}

    \caption{
      Time evolution of the isosurfaces
      $|\Phi|/\eta=0.3$ (opaque white) and
      $|\sigma|/\eta=0.6$ (translucent red),
      computed using the parameters of Set 1 listed in
      Table~\ref{tab:parameters}, with
      $\lambda_{\Phi\sigma}=0.00$, $0.40$, and $0.49$
      from top to bottom.
    }
    \label{fig:TE_strong}
\end{figure}

In Fig.~\ref{fig:TE_strong}, we show the snapshots of the isosurfaces with $|\Phi|/\eta = 0.3$ (white) 
and $|\sigma|/\eta = 0.6$ (red) for the Set 1 simulations listed in Table~\ref{tab:parameters}.
From top to bottom, $\lambda_{\Phi\sigma} = 0.00, 0.40,$ and $0.49$.

After some evolution, cosmic strings form. 
For $\lambda_{\Phi\sigma}=0.49$, the current-carrier field $\sigma$ has a large amplitude in the whole computational domain.
Once the fields have relaxed, the current-carrier field $\sigma$ vanishes far from the strings, while it maintains a large value near the string cores (middle panels). By the end of the simulations, superconducting cosmic string networks have clearly formed. The condensation of the current-carrier field around the string cores is highly sensitive to the parameter $\lambda_{\Phi\sigma}$. In fact, the current-carrier field $\sigma$ is distributed almost uniformly around the string cores for $\lambda_{\Phi\sigma}=0.49$, but becomes remarkably inhomogeneous along strings when $\lambda_{\Phi\sigma}$ is decreased to $0.40$.

%-----------------------------------------------------------%
\subsection{Condensate, charge, and current}
%-----------------------------------------------------------%
To investigate physical quantities around superconducting cosmic strings, we introduce the following weighted average:
\begin{align}
\label{weight_av}
\langle X \rangle \equiv \frac{\displaystyle \int_V d^3x \;X(\boldsymbol{x})(|\Phi/\eta|^p-1)^2}{\displaystyle \int_V d^3x \;(|\Phi/\eta|^p-1)^2} \quad, \quad(p>0).
\end{align}
Here, the weight $(|\Phi/\eta|^p-1)^2$ approaches unity near the core of a cosmic string and vanishes far away from it. Owing to this property, one can extract typical values of physical quantities around the string core. The parameter $p$ is a positive real number, and smaller values of $p$ correspond to a broader weight profile. 
In particular, we fix $p=2$ because the weight is proportional to the potential of the string-forming field, $(|\Phi|^2-\eta^2)^2$, giving a clear physical meaning to the weighted average. Although the numerical values depend on $p$, the overall qualitative time evolution remains similar, with larger $p$ showing larger deviations from the exact $\tau^{-2}$ scaling (see Appendix~\ref{sec:wei_av} for details).
The quantity $X(\boldsymbol{x})$ denotes the physical observable to be averaged, and in this paper we consider the following three quantities:
\begin{align}
X(\boldsymbol{x}) = {|\sigma|^2},\;Q^2,\;\text{and}\;J^2.
\end{align}
Here, $Q^2$ and $J^2$ denote the squared charge and current measures defined from the Noether current of $\sigma$, $j_\mu \equiv 2\text{Im}(\sigma^*\partial_\mu\sigma)$, as
\begin{align}
Q^2 \equiv a^{-2}(j_0)^2,\qquad J^2\equiv a^{-2}\delta^{ik}j_ij_k\;.
\end{align}

\begin{figure*}[htbp]
    \centering
        \includegraphics[width=0.45\linewidth]{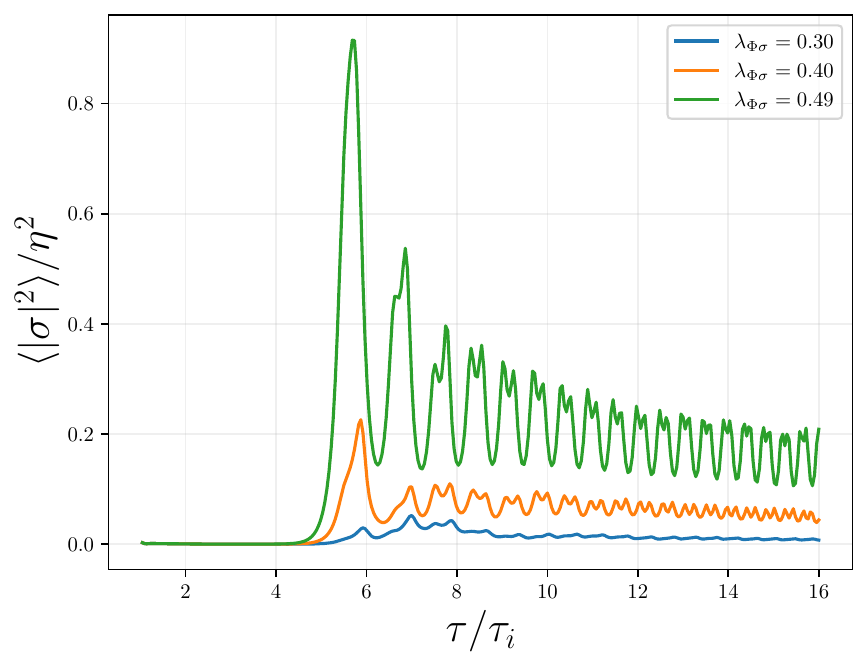}
            \label{fig:wn_condensity_all}
    \caption{Time evolution of the weighted condensate $\langle|\sigma|^2\rangle$.}
    \label{fig:condensity_all}
\end{figure*}

\begin{figure*}[htbp]
    \centering
    \begin{tabular}{c@{\hspace{0.5cm}}c}
        \begin{minipage}{0.45\linewidth}
            \centering
            \includegraphics[width=\linewidth]{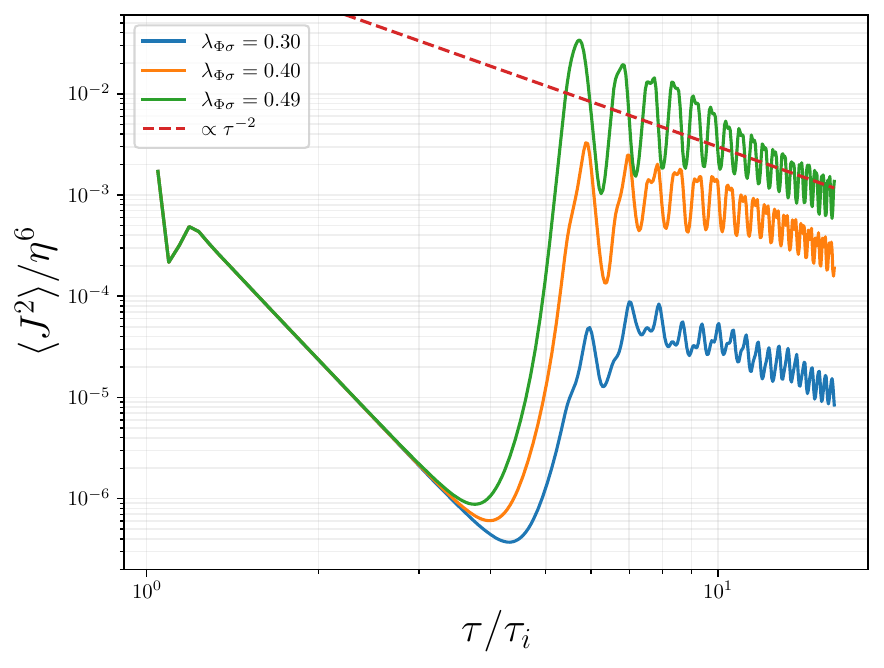}
        \end{minipage}
        &
        \begin{minipage}{0.45\linewidth}
            \centering
            \includegraphics[width=\linewidth]{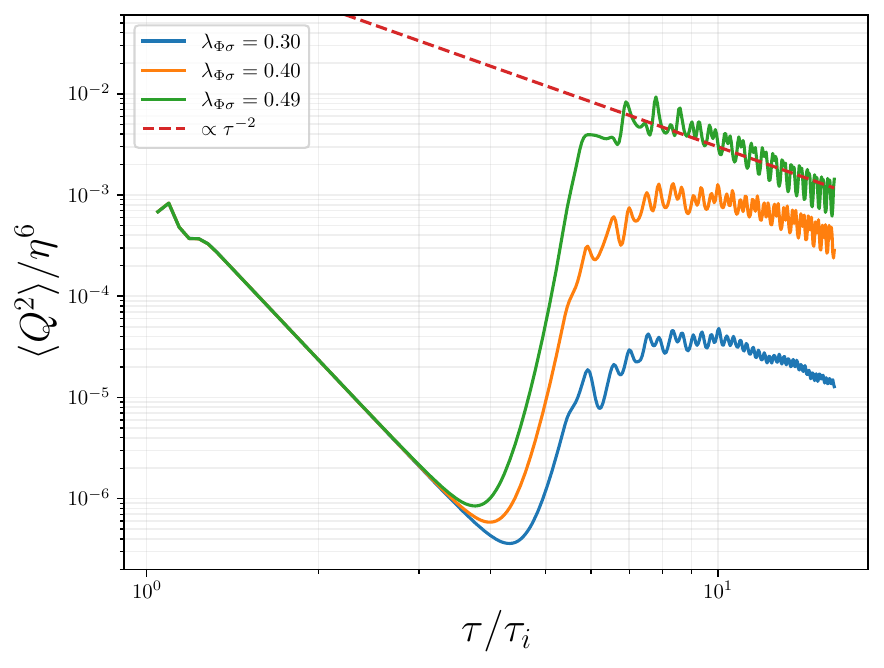}
        \end{minipage}
    \end{tabular}
    \caption{Time evolution of the weighted averages of the squared current and charge.}
    \label{fig:current-charge}
\end{figure*}

Figures~\ref{fig:condensity_all} and \ref{fig:current-charge} show the time evolution of the weighted averages
$\langle|\sigma|^2\rangle$, $\langle Q^2\rangle$, and
$\langle J^2\rangle$ for different values of $\lambda_{\Phi\sigma}$.
All three weighted quantities increase with $\lambda_{\Phi\sigma}$.
Figure~\ref{fig:condensity_all} shows that, after its initial growth,
the weighted average of $|\sigma|^2$ decreases and tends toward an
approximately constant value at late times, most clearly for
$\lambda_{\Phi\sigma}=0.49$. This behavior suggests that the current-carrier field initially excited away from the cosmic strings gradually relaxes,
while the condensate localized around the string cores remains.
Figure~\ref{fig:current-charge} shows that the weighted averages of $Q^2$ and $J^2$ decrease approximately as $\tau^{-2}$. 
A qualitatively similar decrease in the charge and current measures was reported in Ref.~\cite{Correia:2024wsq}, although their observables are defined differently and their simulations employ the fat-string prescription.

A possible interpretation of the relation between the charge and current amplitudes is provided by the dynamics of the current-carrier phase. 
Expressing the current-carrier field as $\sigma=|\sigma|e^{i\theta}$, its Noether current is given by $j_\mu=2|\sigma|^2\partial_\mu\theta$.
For a locally straight superconducting string aligned with the $z$-axis, the coarse-grained phase configuration is conventionally parametrized as
$\theta \sim kz-\omega\tau$~\cite{Hiramatsu:2023epr}.
This configuration may be regarded as a coarse-grained description of the Kibble--Zurek mechanism, analogous to the experiment proposed by Zurek~\cite{Zurek:1985qw}. 
After the phase transition, causally disconnected string segments generally select different phases of the current carrier field, and the resulting current is determined by the phase variation along the string. 
The corresponding correlation length is expected to be of order the horizon scale, with $k$ characterizing its inverse.

In addition to this coarse-grained configuration, the spontaneous breaking of $U(1)_{\rm global}$ in the string core gives rise to approximately massless phase excitations propagating along the string. At the linear level, a general fluctuation can be decomposed into left- and right-moving modes,
\begin{align}
\label{eq:phase_fluctuation}
\theta\sim kz-\omega\tau +\theta_{+}(z-\tau)+\theta_{-}(z+\tau).
\end{align}
Assuming that $\theta_{\pm}$ describe oscillatory fluctuations around zero, the quadratic current and charge measures along the string can be estimated as
\begin{align}
\left\langle j_z^2\right\rangle_{\rm str}
&\simeq 4|\sigma|^4
\left(k^2+\langle\theta_{+}'^2\rangle_{\rm str}+\langle\theta_{-}'^2\rangle_{\rm str}+2\langle\theta_{+}'\theta_{-}'\rangle_{\rm str}\right), \label{eq:jz_str}
\\
\left\langle j_0^2\right\rangle_{\rm str}
&\simeq 4|\sigma|^4\left(\omega^2+\langle\theta_{+}'^2\rangle_{\rm str}+\langle\theta_{-}'^2\rangle_{\rm str}-2\langle\theta_{+}'\theta_{-}'\rangle_{\rm str}\right),\label{eq:j0_str}
\end{align}
where $\langle\cdots\rangle_{\rm str}$ denotes an average along the string. 
If the correlation of the left- and right-moving components can be neglected, $\langle\theta_{+}'\theta_{-}'\rangle_{\rm str}\simeq0$, and their contribution dominates over the background terms, $\langle\theta_{\pm}'^2\rangle_{\rm str}\gtrsim k^2$ and $\langle\theta_{\pm}'^2\rangle_{\rm str}\gtrsim\omega^2$, $\left\langle j_z^2\right\rangle_{\rm str}$ and $\left\langle j_0^2\right\rangle_{\rm str}$ become comparable. 
This provides a possible interpretation of
the similar late-time magnitudes of $\langle Q^2\rangle$ and
$\langle J^2\rangle$ shown in Fig.~\ref{fig:current-charge}.

Equations (\ref{eq:jz_str}) and (\ref{eq:j0_str}) may also explain the approximate $\tau^{-2}$ decay of $\langle Q^2 \rangle$ and $\langle J^2 \rangle$ observed at late times.
For freely propagating massless phase modes confined to the string, cosmic expansion does not distort their profiles in comoving coordinates. Consequently, the statistics of their comoving phase gradients, such as $\langle\theta_\pm^{\prime 2}\rangle_{\rm str}$, remain approximately unchanged during time evolution.
Provided that the radial profile also approximately unchange, the explicit factor $a^{-2}$ in the definitions of $Q^2$ and $J^2$ then gives
\begin{align}
\left\langle Q^2\right\rangle,
\left\langle J^2\right\rangle
\propto a^{-2}.
\end{align}
Since $a\propto\tau$ in the radiation-dominated background considered here, this gives the approximate behavior
$\langle Q^2\rangle,\langle J^2\rangle\propto\tau^{-2}$ observed in Fig.~\ref{fig:current-charge}.

%-----------------------------------------------------------%

%-----------------------------------------------------------%
\subsection{Gravitational waves}\label{subsec:GW}
%-----------------------------------------------------------%

We evaluate the power spectrum of the stochastic gravitational-wave background~\cite{Maggiore:1999vm},
\begin{align}
\Omega_{\rm GW} = \frac{1}{\rho_c}\frac{d\rho_{\rm GW}}{d\log k},
\end{align}
where $\rho_{\rm GW}$ and $\rho_c$ are the gravitational-wave energy density (see Appendix A for details of the calculation) and the critical density of the Universe, respectively.

\begin{figure}[htbp]
    \centering
            \includegraphics[width=.65\linewidth]{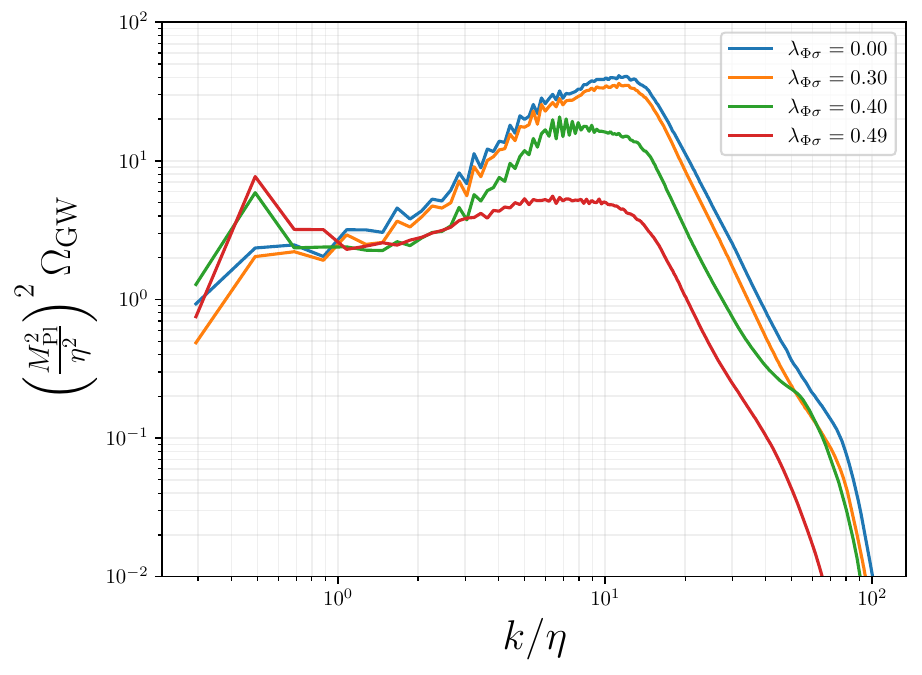} 
            \label{fig:omegagw_wn}
    \caption{Gravitational-wave power spectra with $\lambda_{\Phi\sigma} = 0.00,0.30,0.40,\;\text{and}\;0.49$. Each spectrum is obtained from a single realization.}
    \label{fig:omegagw}
\end{figure}

\begin{figure}[htbp]
    \centering
            \includegraphics[width=.65\linewidth]{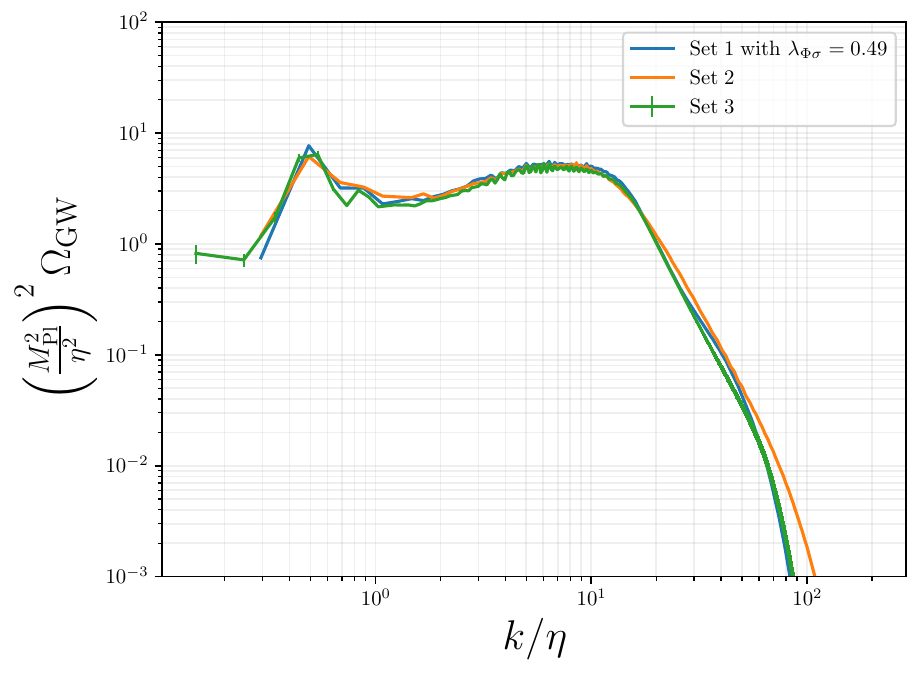} 
    \caption{Gravitational-wave power spectra computed for the same physical setup. The spectrum for Set 1 shown in this figure is identical to that for $\lambda_{\Phi\sigma}=0.49$ shown in Fig.~\ref{fig:omegagw}. For Sets 1 and 2, each spectrum is obtained from a single realization, whereas the Set 3 spectrum is averaged over five realizations with different initial conditions, shown with the error bars.} 
    \label{fig:Omegaboxdxdep}
\end{figure}

Figure \ref{fig:omegagw} shows the dependence of the gravitational-wave power spectrum on the coupling constant $\lambda_{\Phi\sigma}$ between the string-forming and current-carrier fields.
As $\lambda_{\Phi\sigma}$ increases, the power spectrum decreases on small scales but increases on large scales. 
To confirm that this tendency is not a numerical artifact, we perform simulations with different box sizes and lattice spacings while keeping the physical setup unchanged, as noted in Sec.~III. 
Figure \ref{fig:Omegaboxdxdep} shows  the resulting power spectra for the parameter sets listed in Table~\ref{tab:parameters}, confirming that the trend is robust. In particular, the error bars on the averaged spectrum obtained from Set 3 are relatively small compared with the spectral amplitude, indicating limited statistical uncertainty. Moreover, the averaged spectrum is consistent with that obtained from a single realization of Set 1. These results suggest that the main spectral features discussed below can be captured by a single realization. We therefore use a representative single realization to investigate the physical origin of these features.

To understand the behavior, we decompose the energy-momentum tensor (\ref{eq:relavantEMT}) as follows and examine in detail the respective contributions of the cosmic-string and current-carrier components to the gravitational-wave spectrum
\begin{align}
    T_{ij}^{\text{eff}} &= T_{ij}^{\text{string}}+T_{ij}^{\text{cc}},\\
    T_{ij}^{\text{string}}&= 2\text{Re}[(D_{i}\Phi)^* D_{j}\Phi]  + F_{i\mu}F_{j}{}^\mu,\\
    T_{ij}^{\text{cc}}&=2\text{Re}[(\partial_{i}\sigma)^* \partial_{j}\sigma].
\end{align}
In particular, $T_{ij}^{\text{string}}$ is identical to $T_{ij}^{\text{eff}}$ in the Abelian-Higgs model, and the gravitational waves sourced by it can be interpreted as those generated by cosmic strings.

\begin{figure}[htbp]

            \includegraphics[width=0.65\linewidth]{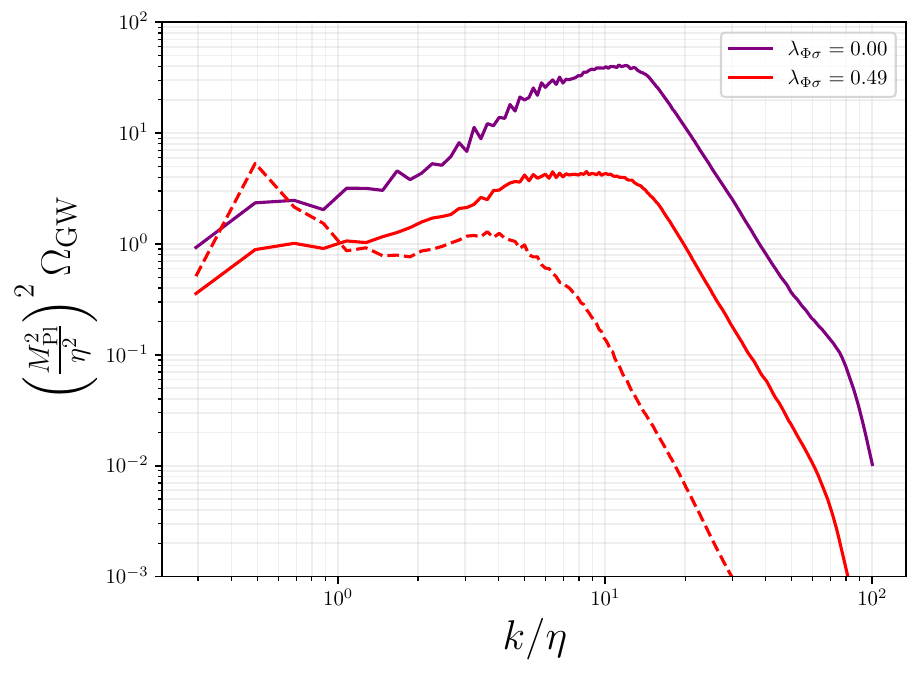}

    \caption{
    Gravitational-wave power spectra sourced by $T_{ij}^{\text{string}}$ (solid lines) and by $T_{ij}^{\text{cc}}$ (dashed lines), computed for $\lambda_{\Phi\sigma}=0.00\;\text{and}\;0.49$
    using the parameters of Set 1 in Table~\ref{tab:parameters}. Each spectrum is obtained from a single realization.}
    \label{fig:omegagw_part_all}
\end{figure}

\begin{figure}[htbp]

\includegraphics[width=0.65\linewidth]{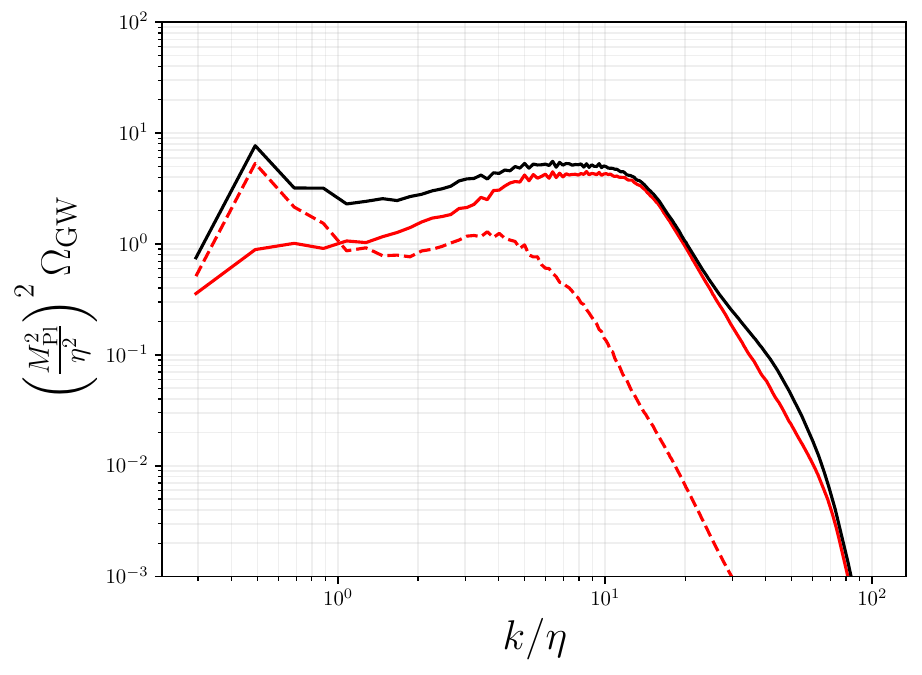}
            \caption{Gravitational-wave power spectra sourced by $T_{ij}^{\text{string}}$ (red solid line), $T_{ij}^{\text{cc}}$ (red dashed line), and $T_{ij}^{\text{eff}}$ (black solid line) for $\lambda_{\Phi\sigma}=0.49$. Each spectrum is obtained from a single realization.}
            \label{fig:omegagw_sum}
\end{figure}

Figure \ref{fig:omegagw_part_all} shows the gravitational-wave spectra computed separately using $T_{ij}^{\text{string}}$ and $T_{ij}^{\text{cc}}$ as sources. For $\lambda_{\Phi\sigma}=0.49$, the large-scale spectrum sourced by $T_{ij}^{\rm cc}$ exceeds that sourced by $T_{ij}^{\rm string}$. It is also found that the spectrum from $T_{ij}^{\text{string}}$ is suppressed over the entire wavenumber range. Figure \ref{fig:omegagw_sum} compares the power spectra sourced by $T_{ij}^{\text{string}}$ and $T_{ij}^{\text{cc}}$, the same ones shown in red solid and dashed lines in Fig.~\ref{fig:omegagw_part_all}, with that sourced by $T_{ij}^{\text{eff}}$. We find that, in the total power spectrum, the contribution from the string-forming field is dominant on small scales, whereas that from the current-carrier field is larger on large scales. 

The suppression of gravitational-waves from the string-forming field $T_{ij}^{\text{string}}$ for $\lambda_{\Phi\sigma}=0.49$ may be understood as follows.
The potential in the model, Eq.~(\ref{potential}), can be rewritten as
\begin{align}
    U(\Phi,\sigma) = \frac{\lambda_\Phi}{4}|\Phi|^4-\frac{\lambda_\Phi}{2}{\left(\eta^2-\frac{2\lambda_{\Phi\sigma}}{\lambda_\Phi}|\sigma|^2\right)}|\Phi|^2 + \frac{m_\sigma^2}
{2}|\sigma|^2 + \frac{\lambda_\sigma}{4}|\sigma|^4 + \text{const}.
\end{align}
Here, the quantity in parentheses in the second term can be interpreted as an effective symmetry-breaking scale, $\eta_\text{eff}^2$. As the current-carrier field condenses more strongly around the strings, $|\sigma|^2$ increases, and hence $\eta_\text{eff}^2$ becomes smaller. Since the symmetry-breaking scale sets a characteristic physical scale of the string-forming sector, its effective reduction may contribute to the suppression of the gravitational-wave spectrum sourced by $T_{ij}^{\text{string}}$, as seen in Fig.~\ref{fig:omegagw_part_all}. 

Another possible mechanism that may contribute to the suppression of gravitational waves is particle emission from the current-carrier field sector. Since the current-carrier field is lighter than the string-forming field in the simulation, particle emission from the current-carrier field can become kinematically accessible at lower frequencies than particle emission from the string-forming field. Cosmic strings may therefore possess an additional energy-loss channel through the emission of current-carrier particles, which could contribute to the suppression of high-frequency gravitational radiation.

From a macroscopic perspective, the modification of the gravitational-wave spectrum may also be interpreted in terms of the equation of state of the superconducting string. For superconducting cosmic strings, the energy per unit length $U$ and the tension $T$ are generally distinct quantities. In the same $U(1)_{\rm local}\times U(1)_{\rm global}$ model as that used in this paper, it has been shown that $U$ increases while $T$ decreases as the current increases~\cite{Peter:1992dw}. Since the transverse propagation speed is given by $c_T^2=T/U$~\cite{Carter:1989xk}, such a change in the equation of state reduces the characteristic speed of transverse perturbations. This may suppress the development of small-scale structure on the strings and consequently reduce gravitational-wave emission at high frequencies.

\begin{figure}[htbp]
            \includegraphics[width=0.6\linewidth]{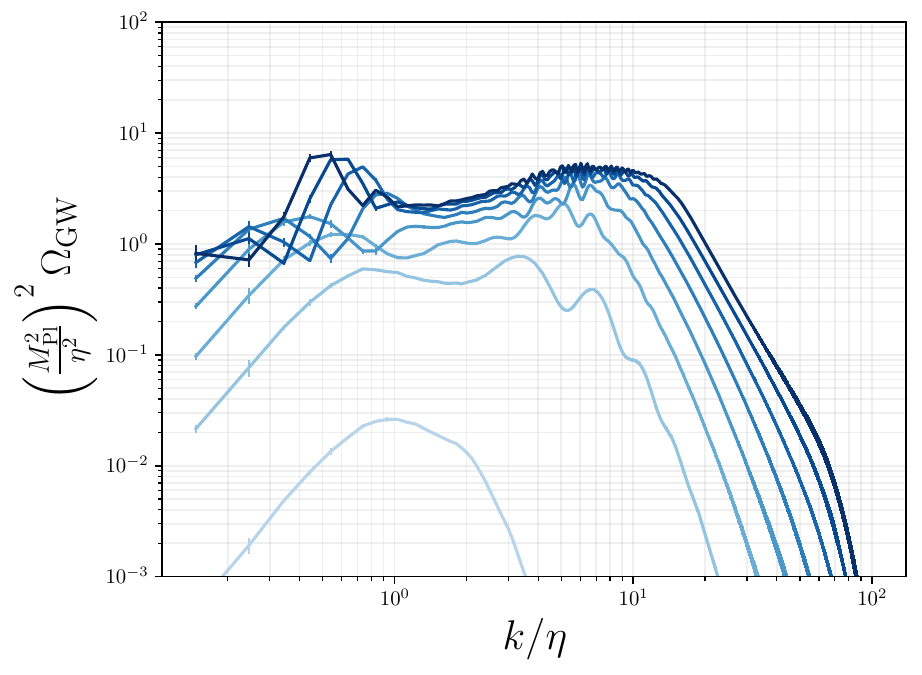}
            \label{fig:mean_evol}
    \caption{
Time evolution of the gravitational-wave power spectrum sourced by $T_{ij}^{\rm eff}$ in Set 3 with $\lambda_{\Phi\sigma}=0.49$. The spectrum is averaged over five realizations with different initial conditions.}
    \label{fig:evol_49_mean-bin}
\end{figure}

Figure~\ref{fig:evol_49_mean-bin} shows the time evolution of the gravitational-wave power spectrum sourced by $T_{ij}^{\rm eff}$. As the system evolves, the low-wavenumber peak shifts toward smaller wavenumbers, while an approximately flat spectral tail remains at higher wavenumbers.

%======================================================================%
\section{Conclusion and discussion}
%======================================================================%
In this paper, we have studied the evolution of a superconducting cosmic string network in a field-theoretic model with $U(1)_{\text{local}}\times U(1)_{\text{global}}$ symmetry. We compute the gravitational-wave spectra from a superconducting cosmic-string network by means of lattice field simulations for the first time. We found that the current-carrier field $\sigma$ condenses around cosmic strings, accompanied by the development of current and charge along them. 

We also showed that the shape of the gravitational-wave power spectrum changes as the coupling constant between the string-forming field, $\Phi$, and the current-carrier field, $\sigma$, varies. In particular, we found that the power spectrum is suppressed on small scales and enhanced on large scales as the coupling constant increases. This observation led us to investigate the respective contributions of cosmic strings and the current-carrier to the gravitational waves. The distinct responses of these components to the interaction strength account for the modification of the total power spectrum. In particular, the large-scale modification is driven by the current-carrier component, whose contribution exceeds that of the cosmic strings. Conversely, the small-scale modification is driven by the cosmic-string component, whose spectrum is suppressed over the entire wavenumber range simulated.

\begin{figure}[htbp]
    \includegraphics[width=.65\linewidth]{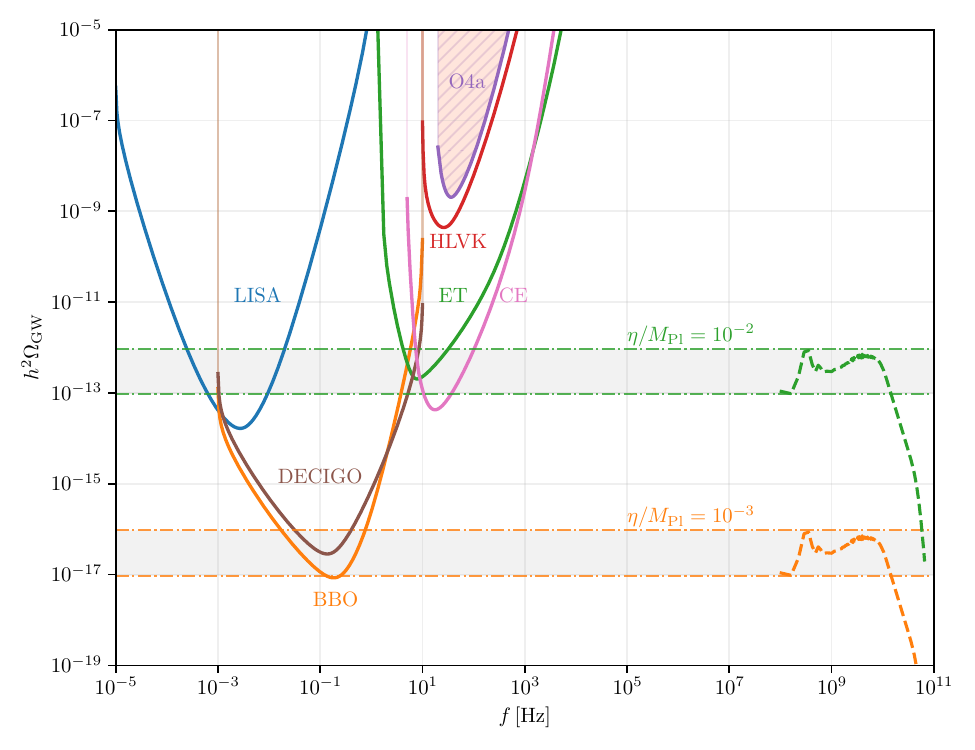}
    \caption{Each region bounded by a pair of dot-dashed horizontal lines of the same color indicates an amplitude range set by the minimum and maximum values of the numerical spectrum shown in Fig.~\ref{fig:evol_49_mean-bin}. The range is evaluated on scales larger than the comoving string width at the final simulation time, $\delta \sim a_{\rm f}^{-1}\eta^{-1}$, corresponding to $k\lesssim \delta^{-1}\sim 10\eta$, and is rescaled using Eq.~(\ref{eq:present_omega}). The corresponding rescaled numerical spectrum is shown on the right for reference as a dashed line, redshifted using Eq.~(\ref{eq:redshift}). The HLV O1--O4a observational constraint is shown as a shaded region~\cite{Virgo:2025aai}, while the sensitivities of future detectors are shown as power-law-integrated sensitivity curves~\cite{Schmitz:2020syl}.
}
    \label{fig:obsdata}
\end{figure}

We now discuss the detectability of gravitational-wave signals from superconducting cosmic strings in current and future observations. To relate the simulated gravitational-wave power spectrum to its present-day
amplitude, we use
\begin{align}
h^2\Omega_{\rm GW,0} \simeq h^2\Omega_{\text{rad},0}\left( \frac{g_{0}}{g} \right)^{1/3}\Omega_{\text{GW}}\simeq
1.34\times10^{-5}
\left(\frac{g}{100}\right)^{-1/3}
\Omega_{\rm GW},
\label{eq:present_omega}
\end{align}
where $\Omega_{\rm GW}$ is the spectrum obtained from the simulations,
$g_0\simeq3.36$ and $g$ are the effective numbers of
relativistic degrees of freedom at the present epoch and at the time of
gravitational-wave production, respectively, and we have used
$h^2\Omega_{\rm rad,0}\simeq4.15\times10^{-5}$.

The present-day frequency of gravitational waves can be estimated by redshifting their physical wavenumber at the final time of the simulation,
\begin{align}
\label{eq:redshift}
f_{\text{phys},0}
&= \frac{k_{\text{phys},\text{f}}}{2\pi}\frac{a_\text{f}}{a_0}= 3.05\times10^{9}\;\text{Hz}\left(\frac{k}{\eta}\right)\left(\frac{g}{100}\right)^{-1/3}\left(\frac{\eta}{T_{\rm i}}\right),
\end{align}
where $k_{\text{phys},\text{f}}\equiv k/a_{\rm f}$ is the physical wavenumber at the final time, $a_\text{f}$ and $a_0$ are the scale factors at the final time of the simulation and at the present time, respectively, and $T_{\rm i}$ is the temperature at the beginning of the simulation, which is $T_{\rm i}\sim \eta$.

Figure~\ref{fig:obsdata} shows the gravitational-wave power spectrum obtained from simulations with the Set 3 parameters, rescaled using Eqs.~(\ref{eq:present_omega}) and (\ref{eq:redshift}) for two values of $\eta/M_{\rm Pl}$ with $g\sim10^2$. Before rescaling, this spectrum is identical to the final-time spectrum shown in Fig.~\ref{fig:evol_49_mean-bin}. The figure also includes the design sensitivity curves of LISA~\cite{Folkner:1998eni,LISA:2017pwj}, DECIGO~\cite{Kawamura:2006up}, BBO~\cite{Corbin:2005ny}, ET~\cite{Punturo:2010zz}, CE~\cite{Reitze:2019iox}, and HLVK~\cite{KAGRA:2013rdx}.

For superconducting cosmic strings in a $U(1)_{\mathrm{local}}\times U(1)_{\mathrm{global}}$ model, Ref.~\cite{Correia:2024wsq} showed that the string network enters the scaling regime, in which its characteristic physical length scale remains proportional to the horizon size, corresponding to an approximately constant number of long strings of order unity per Hubble volume. Although the study in Ref.~\cite{Correia:2024wsq} employed the fat-string regime and we have not quantitatively established whether the string network in our simulations has reached the scaling regime, we may expect superconducting cosmic-string networks to enter the scaling regime in the physical-string regime as well, given that Ref.~\cite{Hiramatsu:2013tga} demonstrated that cosmic-string networks reach scaling in both the fat-string and physical-string regimes, even in the presence of inter-string interactions.

Once the cosmic-string network enters the scaling regime during the radiation-dominated era, the amplitude of the gravitational-wave power spectrum generated by the network is expected to be approximately scale independent for modes generated during the regime.\footnote{Changes in the effective number of relativistic degrees of freedom during cosmic evolution can, however, lead to departures from exact scale invariance in the present-day spectrum; see, e.g., Ref.~\cite{Caldwell:1996en}.} If the simulated superconducting cosmic-string network has reached the scaling regime and continues to exhibit scaling behavior, the amplitude inferred from the numerical spectrum is expected to extend toward lower frequencies. As illustrated in Fig.~\ref{fig:obsdata}, this suggests that future gravitational-wave detectors may probe superconducting cosmic-string networks with $\eta \gtrsim 10^{-3}M_{\rm Pl}$.

As the overall amplitude of $\Omega_{\mathrm{GW}}$ for small $k$ does not deviate significantly from that of the standard Abelian-Higgs string network ($\lambda_{\Phi\sigma} = 0$), as shown in Fig.~\ref{fig:omegagw}, distinguishing superconducting cosmic strings from Abelian-Higgs cosmic strings with current and proposed interferometers remains extremely challenging. 
However, a significant modification to the gravitational-wave power spectrum appears in the high-frequency regime corresponding to the string width or the symmetry-breaking scale, $k\sim a_{\rm f}\eta\sim 10\eta$. 
Figure \ref{fig:omegagw} demonstrates that a larger coupling $\lambda_{\Phi\sigma}$ leads to greater suppression of the gravitational-wave power spectrum at this scale. 
Consequently, future high-frequency gravitational-wave (HFGW) detectors \cite{Aggarwal:2025noe} may clarify whether the observed gravitational waves originate from Abelian-Higgs strings or superconducting ones. Although sensitivity in the MHz-to-GHz range remains challenging today, ongoing advances and novel detector concepts for HFGWs are expected to open a new window into probing the fine structure of cosmic string networks.

\begin{acknowledgments}
The authors would like to thank the anonymous referee for their careful reading and helpful suggestions to improve this paper. We thank Tsutomu Kobayashi for helpful comments on the manuscript, as well as the organizers and participants of Summer Institute 2025 for valuable discussions and comments. The numerical calculations were performed using Yukawa-21 and Heian at the Yukawa Institute for Theoretical Physics (YITP), Kyoto University. JJ was supported by the Rikkyo University Special Fund for Research and the Tonen International Scholarship Foundation. TH was supported by JSPS KAKENHI Grant Numbers JP25K07327 and JP23H00100.
\end{acknowledgments}

\appendix
\section{Dependence of the Weighted-Averaged Quantities on $p$}\label{sec:wei_av}

We used the weighted average defined in Eq.~(\ref{weight_av}) to evaluate the quantities around the strings. Since the spatial extent of the weight function is controlled by the parameter $p$, the resulting time evolution of the condensate, charge, and current may depend on the choice of $p$.
\begin{figure*}[htbp]
    \centering
    \begin{tabular}{c@{\hspace{0.5cm}}c}
        \begin{minipage}{0.45\linewidth}
            \centering
            \includegraphics[width=\linewidth]{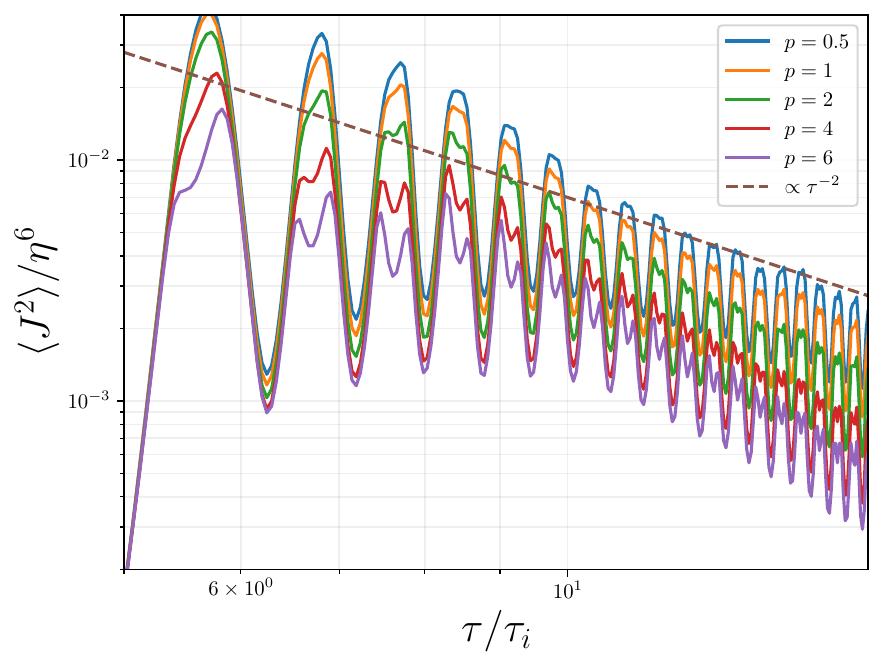}
        \end{minipage}
        &
        \begin{minipage}{0.45\linewidth}
            \centering
            \includegraphics[width=\linewidth]{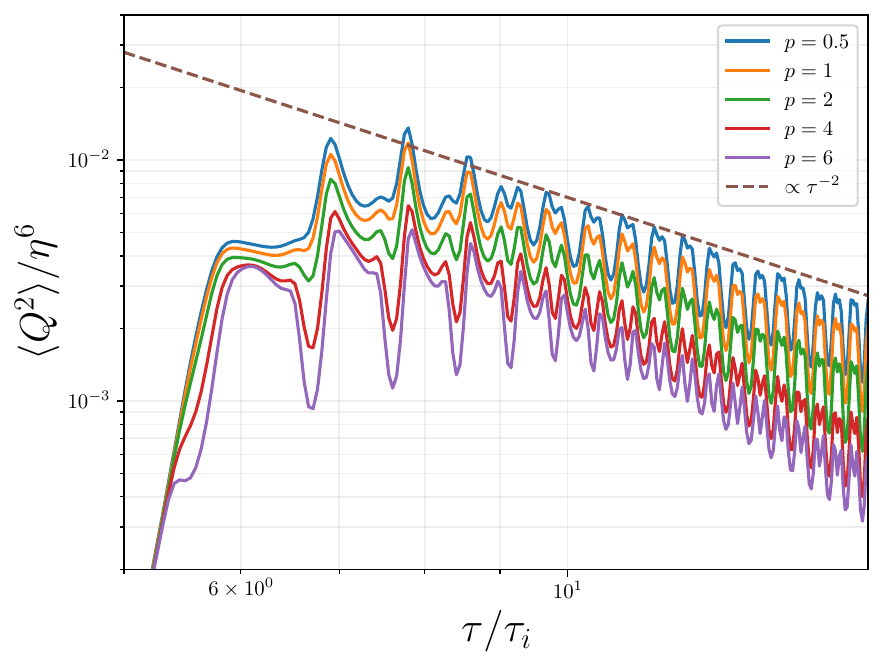}
        \end{minipage}
    \end{tabular}
    \caption{Time evolution of the weighted averages of the squared current and charge with different $p$, using the parameters of Set 1 with $\lambda_{\Phi\sigma} = 0.49$.}
    \label{fig:diff-p-current-charge}
\end{figure*}

Figure~\ref{fig:diff-p-current-charge} shows the $p$ dependence of the weighted-averaged charge and current. As shown in the figure, an approximate $\tau^{-2}$ scaling is observed for all values of $p$ considered here, indicating that the overall time dependence is insensitive to the choice of the weight function. However, the deviation from exact $\tau^{-2}$ scaling becomes more apperent as $p$ increases. This behavior can be understood from the fact that a larger $p$ makes the weight function more localized around the string core. If the weight becomes too narrow compared with the finite width of the string, it may fail to fully reflect the time evolution of the charge and current distributions across the string width.

\begin{figure*}[htbp]
    \centering
        \includegraphics[width=0.45\linewidth]{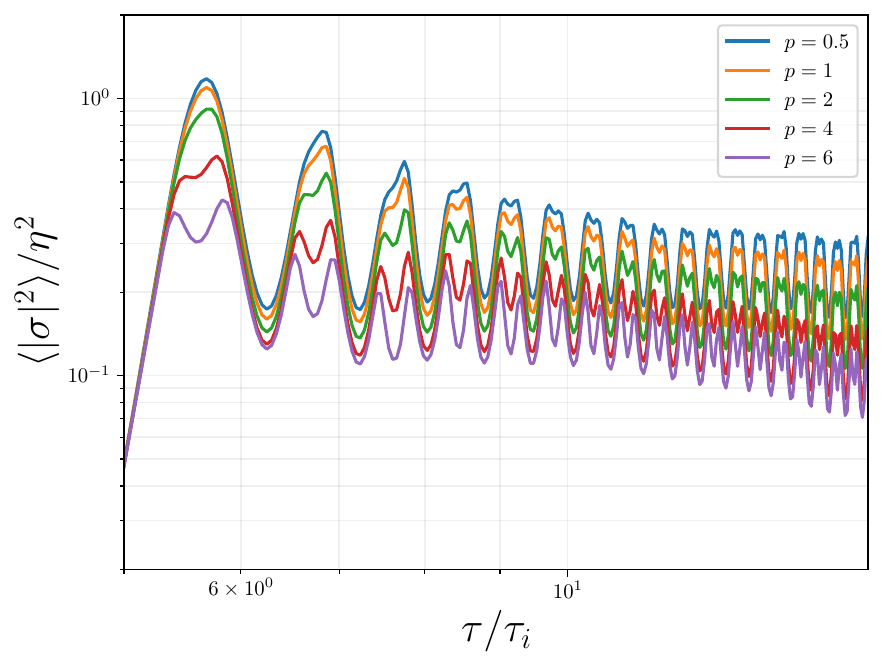}
    \caption{Time evolution of the weighted condensate $\langle|\sigma|^2\rangle$, using the parameters of Set 1 with $\lambda_{\Phi\sigma} = 0.49$.}
    \label{fig:condensity_diff_p}
\end{figure*}

As can be seen from Eq.~(\ref{eq:jz_str}) and (\ref{eq:j0_str}), the estimated $\langle Q^2\rangle$ and $\langle J^2\rangle$ contain a factor proportional to $|\sigma|^4$. This $|\sigma|^4$ dependence makes the estimates of $\langle Q^2\rangle$ and $\langle J^2\rangle$ more sensitive to the choice of $p$. Indeed, Figure~\ref{fig:condensity_diff_p} shows that the weighted-averaged condensate $\langle |\sigma|^2\rangle$ tends to deviate more strongly from a constant value during the time evolution as $p$ increases. This behavior may contribute to the increasing deviation from the $\tau^{-2}$ scaling as $p$ increases in Figure~\ref{fig:diff-p-current-charge}.

\section{Calculation of the Gravitational-Wave Power Spectrum}\label{sec:GWpower}

The energy density of gravitational waves is estimated by
\begin{align}
    \label{GWdensity}
    \rho_{\text{GW}}(\tau) \equiv \frac{1}{32\pi Ga^2}\left\langle h^{\text{TT}}_{ij}{}'(\boldsymbol{x},\tau)h^{\text{TT}}_{ij}{}'(\boldsymbol{x},\tau) \right\rangle_V = \frac{M_{\rm Pl}^{2}}{4a^2 V}\sum_{\boldsymbol{k}}h^{\text{TT}}_{ij}{}'(\boldsymbol{k},\tau)h^{\text{TT}*}_{ij}{}'(\boldsymbol{k},\tau) ,
\end{align}
where $\left\langle \cdots \right\rangle_V$ is the volume average of the simulation box, 
$h^{\text{TT}}_{ij}(\boldsymbol{x},\tau)$ is the transverse-traceless (TT) part of the metric perturbation and $h^{\text{TT}}_{ij}(\boldsymbol{k},\tau)$ is its Fourier component. Employing the discrete Fourier transform, we obtain
\begin{align}
    h^{\text{TT}}_{ij}{}'(\boldsymbol{x},\tau) = \frac{1}{\sqrt{V}} \sum_{\boldsymbol{k}} h^{\text{TT}}_{ij}{}'(\boldsymbol{k},\tau) e^{i\boldsymbol{k}\cdot\boldsymbol{x}}\;.
\end{align}
The extraction of the transverse-traceless part can be performed with the projection tensor, $h^{\rm TT}_{ij}(\boldsymbol{k},\tau) = \Lambda^{\rm TT}_{ij,k\ell}(\hat{k})h_{k\ell}(\boldsymbol{k},\tau)$, where $\Lambda^{\rm TT}_{ij,k\ell}$ is given by
\begin{align}
    \Lambda^{\rm TT}_{ij,k\ell}(\hat{\boldsymbol{k}}) \equiv P_{i\ell}(\hat{\boldsymbol{k}})P_{jm}(\hat{\boldsymbol{k}}) - \frac{1}{2}P_{ij}(\hat{\boldsymbol{k}})P_{\ell m}(\hat{\boldsymbol{k}})  ,
\end{align}
where $P_{ij}(\hat{\boldsymbol{k}})\equiv \delta_{ij} - \hat{k}_i\hat{k}_j$ and $\hat{k}_i \equiv k_i/\sqrt{k_1^2+k_2^2+k_3^2}$.
Although a TT projection tensor can be constructed more consistently on a lattice by
using lattice momenta, we employ the conventional continuum-based one defined above. Previous studies have shown that this choice gives GW spectra that agree well with those obtained using lattice-based ones,
apart from possible differences in the UV tail \cite{Figueroa:2011ye}.
Therefore, lattice-specific corrections to the TT projection are not considered
in the following.
Assuming that the box is sufficiently large, the power spectrum of the stochastic gravitational waves can be written as follows:
\begin{align}
\label{GWPower}
    \Omega_{\text{GW}} = \frac{k^3}{24\pi^2 \mathcal{H}^2}\int \frac{d\Omega_k}{4\pi} h^{\text{TT}}_{ij}{}'(\boldsymbol{k},\tau)h^{\text{TT}*}_{ij}{}'(\boldsymbol{k},\tau),
\end{align}
where $\int d\Omega_{k}$ denotes the angular integral for a fixed $k$.

\bibliographystyle{JHEP}
\bibliography{ref}

\end{document}